\documentclass{pinchcr}   

\usepackage[latin1]{inputenc}
\usepackage{graphicx,amsmath,amssymb,bm,subfigure,comment,fancybox}
\usepackage{epsfig}
\usepackage{bm}  
\newcommand{\beq}{\begin{eqnarray}}
\newcommand{\eeq}{\end{eqnarray}}
\newcommand\eqn[1]{\label{eq:#1}} 
\newcommand\eq[1]{eqn.~\eqref{eq:#1}} 
\newcommand\eqs[2]{eqns.~\eqref{eq:#1}-\eqref{eq:#2}} 
\newcommand\half{{\textstyle{\frac{1}{2}}}}

\newcommand{\Tr}{\mathop{\rm Tr\,}}
\newcommand\bra[1]{\langle #1 \vert}
\newcommand\ket[1]{\vert #1 \rangle}
\newcommand\braket[2]{\langle #1 \vert #2 \rangle}
\newcommand\expect[3]{\langle #1 \vert #2\vert #3 \rangle}
\newcommand\vev[1]{\langle #1 \rangle}
\newcommand{\sla}[1]{\kern .25em\raise.18ex\hbox{$/$}\kern-.80em #1}
\newcommand{\Dslash}{\hbox{\kern .25em\raise.18ex\hbox{$/$}\kern-.80em  $D$}}
\newcommand{\dslash}{\hbox{\kern .22em\raise.18ex\hbox{$/$}\kern-.60em  $\partial$}}
\newcommand{\pslash}{\hbox{\kern .22em\raise.18ex\hbox{$/$}\kern-.60em  $p$}}
\newcommand{\CA}{{\mathcal  A}}

\newcommand{\CH}{{\mathcal  H}}
\newcommand{\CL}{{\mathcal  L}}

\newcommand{\CO}{{\mathcal  O}}

\newcommand{\CQ}{{\mathcal Q}}
\newcommand{\CR}{{\mathcal R}}

\newcommand{\CU}{{\mathcal U}}

\newcommand{\bfm}{{\bf m}}
\newcommand{\bfn}{{\bf n}}

\newcommand{\bfp}{{\bf p}}
\newcommand{\bfq}{{\bf q}}

\newcommand{\bfx}{{\bf x}}
\newcommand{\mybar}[1]{\kern 0.6pt\overline{\kern -0.6pt#1\kern -0.6pt}\kern 0.6pt}

\title{Chiral Symmetry and Lattice Fermions}

\author{David B. Kaplan}
\affiliation{\vbox{Institute  for Nuclear Theory, University of Washington, Seattle, WA 
    98195-1550\\ \  \\Lectures delivered at the Les Houches \'Ecole  d'\'Et\'e de Physique Th\'eorique}}
\begin{document}

\maketitle




\acknowledgements
I have benefitted enormously in my understanding of lattice fermions from conversations with  Maarten Golterman,  Pilar Hernandez,  Yoshio Kikukawa,  Martin L\"uscher,  Rajamani Narayanan,  Herbert Neuberger, and Stephen Sharpe.  Many thanks  to them (and no blame for the content of these notes).
This work is supported in part by U.S.\ DOE grant
No.\ DE-FG02-00ER41132.

\tableofcontents

\maintext

%
%

\chapter{Chiral symmetry}

\section{Introduction}

Chiral symmetries play an important role in the spectrum and phenomenology of both the standard model and various theories for physics beyond the standard model.  In many cases chiral symmetry is associated with nonperturbative physics which can only be quantitatively explored in full on a lattice.  It is therefore important to implement chiral symmetry on the lattice, which turns out to be less than straightforward.  In these lectures I discuss what chiral symmetry is,  why it is important, how it is broken, and ways to implement it on the lattice.   There have been many hundreds of papers on the subject and this is not an exhaustive review; the limited choice of topics I cover reflects on the scope of my own understanding and not the value of the omitted work.

\section{Spinor representations of the Lorentz group}
\label{Lorentz}

To understand chiral symmetry one must understand Lorentz symmetry first.  Since we will be discussing fermions in various dimensions of spacetime, consider the generalization of the usual Lorentz group to $d$ dimensions.   The Lorentz group 
is defined by the real matrices
  $\Lambda$ which
preserve the form of the $d$-dimensional metric
\beq
\Lambda^T \eta \,\Lambda = \eta\ ,\qquad \eta=\text{diag}\left(1,-1,\ldots-1\right).
\eqn{Llorentz}\eeq
With this definition, the inner product between two 4-vectors, $v^\mu \eta_{\mu\nu} w^\mu =
v^T\eta w$,  is preserved under the Lorentz transformations $v\to \Lambda v$ and $w\to \Lambda w$.  This defines the group $SO(d-1,1)$, which --- like $SO(d)$ --- has $d(d-1)/2$ linearly independent generators, which may be written as $M^{\mu\nu} = - M^{\nu\mu}$, where the indices $\mu,\nu=0,\ldots,(d-1)$ and
\beq
\Lambda = e^{i\theta_{\mu\nu}M^{\mu\nu}}\ ,
\eeq
with $\theta_{\mu\nu} = -\theta_{\nu\mu}$ being $d(d-1)/2$ real parameters.  Note that $\mu,\nu$ label the $d(d-1)/2$ generators, while in a representation $R$ each $M$ is a $d_R\times d_R$ matrix, where $d_R$ is the dimension of $R$.
By expanding \eq{Llorentz} to order $\theta$ one sees
that the generators $M$ must satisfy
\beq
(M^{\mu\nu})^T \eta + \eta M^{\mu\nu} =0.
\eqn{lorentz}\eeq
From this equation it is straightforward to write down a basis for the $M^{\mu\nu} $ in the $d$-dimensional defining representation and determine the commutation relations for the algebra,
\beq
\left[M^{\alpha\beta}, M^{\gamma\delta}\right] = i \left(\eta^{\beta\gamma} M^{\alpha\delta}-\eta^{\alpha\gamma} M^{\beta\delta}-\eta^{\beta\delta} M^{\alpha\gamma}+\eta^{\alpha\delta} M^{\beta\gamma}\right).
\eeq
A Dirac spinor representation can be constructed as
\beq
M^{\alpha\beta} \equiv \Sigma^{\alpha\beta}= \frac{i}{4}\left[\gamma^\alpha,\,\gamma^\beta\right]
\eqn{spinor}\eeq
where the gamma matrices satisfy the Clifford algebra:
\beq
\{\gamma^\alpha,\,\gamma^\beta\} = 2\eta^{\alpha\beta}
\eqn{gamma}
\eeq

Solutions to the Clifford algebra are easy to find by making use of direct products of Pauli matrices. In a direct product space we can write a matrix as $M=a\otimes A$ where $a$ and $A$ are matrices of dimension $d_a$ and $d_A$ respectively, acting in different spaces; the matrix $M$  then has dimension  $(d_a\times d_A)$.  Matrix multiplication is defined as $(a\otimes A)(b\otimes B)= (ab)\otimes (AB)$.  It is usually much easier to construct a representation when you need one rather than to look one up and try to keep the conventions straight!
One finds that solutions for the $\gamma$ matrices in $d$-dimensions obey the following properties:
\begin{enumerate}
\item For both $d=2k$ and $d=2k+1$, the $\gamma$-matrices are $2^k$ dimensional;
\item For even spacetime dimension $d=2k$  (such as our own with $k=2$) one can define a generalization of $\gamma_5$ to be
\beq
\Gamma = i^{k-1} \prod_{\mu=0}^{2k-1} \gamma^\mu
\eqn{Gammadef}
\eeq
with the properties 
\beq
\{\Gamma,\,\gamma^\mu\} = 0\ ,\quad
\Gamma=\Gamma^\dagger=\Gamma^{-1}\ ,\quad \Tr ( \Gamma \gamma^{\alpha_1}\cdots\gamma^{\alpha_{2k}}) =2^k i^{-1-k}\epsilon^{\alpha_1\ldots\alpha_{2k}}\ ,
\eqn{Gammaprop}
\eeq
where $\epsilon_{012\ldots 2k-1}=+1=-\epsilon^{012\ldots 2k-1}$.
\item In $d=2k+1$  dimensions one needs one more $\gamma$-matrix than in $d=2k$, and one can take it to be $\gamma^{2k} = i\Gamma$.
\end{enumerate}
Sometimes it is useful to work in a specific basis for the $\gamma$-matrices; a particulary useful choice is a ``chiral basis", defined to be one where $\Gamma$ is diagonal.  For example, for $d=2$ and $d=4$ (Minkowski spacetime) one can choose

\begin{align}
d=2:&\quad
\gamma^0  =  \sigma_1\ ,\quad
\gamma^1  = - i\sigma_2\ ,\quad
\Gamma = \sigma_3\ \eqn{d2chiral}\\ &\cr
d=4:&
\quad
\gamma^0  = -\sigma_1\otimes 1\ ,\quad
\gamma^i  =   i\sigma_2\otimes \sigma_i\ ,\quad
\Gamma=  \sigma_3\otimes 1\ .
\eqn{d4chiral}%
\end{align}

\subsection{$\gamma$-matrices in Euclidian spacetime}

In going to Euclidian spacetime with metric $\eta^{\mu\nu}=\delta_{\mu\nu}$, one takes 
\beq
\partial_0^M\to i\partial_0^E\ ,\quad \partial_i^M\to \partial_i^E
\eqn{euc1}\eeq
and defines 
\beq
\gamma^0_M= \gamma^0_E\ ,\quad \gamma^i_M= i\gamma^i_E\ ,
\eeq
 so that 
 \beq
 (\gamma_E^\mu)^\dagger=\gamma_E^\mu\ ,\qquad \{\gamma_E^\mu,\gamma_E^\nu\}=2\delta_{\mu\nu}
 \eeq
  and $\Dslash_M\to i\Dslash_E$, with 
  \beq
  \Dslash_E=-\Dslash_E^\dagger
  \eeq
  and the Euclidian Dirac operator is $(\Dslash_E+m)$.
  The  matrix $\Gamma^{(2k)}$ in $2k$ dimensions is taken to equal $\gamma^{2k}_E$ in $(2k+1)$ dimensions:
  \beq
 \Gamma^{(2k)}_E = \gamma^{2k}_E = \Gamma^{(2k)}_M\ ,\qquad\Tr ( \Gamma_E\, \gamma_E^{\alpha_1}\cdots\gamma_E^{\alpha_{2k}}) =-2^k i^{k}\epsilon^{\alpha_1\ldots\alpha_{2k}}
 \eqn{euc2}\eeq
 where $\epsilon_{012\ldots 2k-1}=+1=+\epsilon^{012\ldots 2k-1}$.

\section{Chirality in even dimensions}
\label{Chirality}

My lectures on chiral fermions follow from the properties of the matrix $\Gamma$ in even dimensions.  The existence of $\Gamma$  means that Dirac spinors are reducible representations of the Lorentz group, which in turn means we can have symmetries (``chiral symmetries'') which transform different parts of Dirac spinors in different ways.  To see this,   define the projection operators
\beq
P_\pm = \frac{(1\pm\Gamma)}{2}\ ,
\eeq
which have the properties
\beq
\qquad P_+ + P_- = {\mathbf 1}\ ,\qquad P_\pm^2 = P_\pm\ ,\qquad P_+P_-=0\ .
\eeq
 Since in odd spatial dimensions $\{\Gamma,\gamma^\mu\}=0$ for all $\mu$, it immediately follows that $\Gamma$ commutes with the Lorentz generators $\Sigma^{\mu\nu}$ in \eq{spinor}:  $\left[\Gamma,\Sigma^{\mu\nu}\right]=0$. Therefore we can write $\Sigma^{\mu\nu}= \Sigma_+^{\mu\nu}+\Sigma_-^{\mu\nu}$ where
 \beq
 \Sigma^{\mu\nu}_\pm = P_\pm \Sigma^{\mu\nu}P_\pm\ ,\qquad  \Sigma^{\alpha\beta}_+ \Sigma^{\mu\nu}_-=\Sigma^{\mu\nu}_- \Sigma^{\alpha\beta}_+ =0\ .
 \eeq
 Thus $\Sigma^{\mu\nu}$ is reducible:  spinors $\psi_{\pm}$ which are eigenstates of $\Gamma$ with eigenvalue $\pm 1$ respectively transform independently under Lorentz transformations. 

The word  ``chiral" comes from the Greek word for hand, $\chi\epsilon\iota\rho$.  The projection operators $P_+$ and $P_-$ are often called $P_R$ and $P_L$ respectively; what does handedness have to do with the matrix $\Gamma$?  Consider the Lagrangian for a free massive Dirac fermion in $1+1$ dimensions and use the  definition of $\Gamma=\gamma^0\gamma^1$:
\beq
\CL & =&\mybar\Psi i\dslash\Psi\cr
 &=& \Psi^\dagger i\left(\partial_t +\Gamma\partial_x\right)\Psi\cr
& =&
\Psi_+^\dagger i\left(\partial_t +\partial_x\right)\Psi_+ +  \Psi_-^\dagger i\left(\partial_t -\partial_x\right)\Psi_- \eeq
where
\beq
\Gamma \Psi_\pm = \pm \Psi_\pm\ .
\eeq
%
We see then that the solutions to the Dirac equation for $m=0$ are $$\Psi_\pm(x,t) = \Psi_\pm(x\mp t)$$ so that $\Psi_+$ corresponds to right-moving solutions, and $\Psi_-$ corresponds to left-moving solutions. This is possible since the massless particles move at the speed of light and the direction of motion is invariant under proper Lorentz transformations in $(1+1)$ dimensions.

In the chiral basis \eq{d2chiral},  the positive energy plane wave solutions to the Dirac equation are
\beq
\Psi_+ = e^{-iE(t-x)} \begin{pmatrix} 1\cr 0 \end{pmatrix}\ ,\qquad 
\Psi_- = e^{-iE(t+x)} \begin{pmatrix} 0\cr 1 \end{pmatrix}\ .
\eeq
 It is natural to call $P_+$ and $P_-$ as $P_R$ and $P_L$ respectively.

\begin{exercise}
\label{ex1b}
You should perform the same exercise in $3+1$ dimensions and find that solutions $\Psi_\pm$ to the massless Dirac equation satisfying $\Gamma\Psi_\pm = \pm \Psi_\pm$ must also satisfy $|\vec p|=E$ and $(2\vec p\cdot\vec S/E)  \Psi_\pm =\pm\Psi_\pm$, where $S_i =\half \epsilon_{0ijk }\Sigma^{jk }$ are the generators of rotations.  Thus $\Psi_\pm$ correspond to states with positive or negative helicity respectively, and are called right- and left-handed particles.
\end{exercise}

\section{Chiral symmetry and fermion mass in four dimensions}
\label{chisym}

Consider the Lagrangian for a single flavor of Dirac fermion in 3+1 dimensions, coupled to a background gauge field
\beq
\CL = \mybar\Psi (i\Dslash - m)\Psi  = \left( \mybar\Psi_L i\Dslash \Psi_L +  \mybar\Psi_R i\Dslash \Psi_R\right) -m \left(\mybar\Psi_L\Psi_R +\mybar\Psi_R \Psi_L\right)
\eeq
where I have defined
\beq
\Psi_L = P_-\Psi\ ,\quad \bar\Psi_L = \Psi_L^\dagger\gamma^0 = \bar\Psi P_+\ ,\quad \Psi_R = P_+\Psi\ ,\quad \bar\Psi_R=\bar\Psi P_-\ .
\eqn{dl1}
\eeq
For now I am assuming that $\Psi_{L,R}$ are in the same complex representation of the gauge group, where $D_\mu$ is the gauge covariant derivative appropriate for that representation. It is important to note the the property $\{\gamma_5,\gamma^\mu\}=0$ ensured that the kinetic terms in \eq{dl1} do not couple left-handed and right-handed fermions; on the other hand, the mass terms do\footnote{I will use the familiar $\gamma_5$ in $3+1$ dimensions instead of $\Gamma$  when there is no risk of ambiguity.}. The above Lagrangian has an exact $U(1)$ symmetry, associated with fermion number, $\Psi\rightarrow e^{i\alpha}\Psi$.  Under this symmetry, left-handed and right-handed components of $\Psi$ rotate with the same phase; this is often called a ``vector symmetry''.   In the case where $m=0$, it apparently has an additional symmetry where the left- and right-handed components rotate with the opposite phase, $\Psi\rightarrow e^{i\alpha \gamma_5}\Psi$; this is called an ``axial symmetry'', $U(1)_A$. 

Symmetries are associated with Noether currents, and symmetry violation appears as a nonzero divergence for the current.  Recall the Noether formula for a field $\phi$ and infinitesimal transformation $\phi \to \phi +  \epsilon\delta\phi$:
\beq
J^\mu =-\frac{\partial \CL}{\partial(\partial_\mu\phi)}\delta\phi\ ,\qquad \partial_\mu J^\mu = -\delta\CL\ .
\eeq

In the Dirac theory, the vector symmetry corresponds to $\delta\Psi = i \Psi$, and the axial symmetry transformation is $\delta\Psi = i \gamma_5\Psi$, so that the
  Noether formula yields the vector and axial currents:
\begin{align}
U(1)&:\qquad J^\mu = \mybar\Psi\gamma^\mu \Psi \ , &\partial_\mu J^\mu = &0&&\\
U(1)_{A} &:\qquad J^\mu_A =  \mybar\Psi\gamma^\mu \gamma_5\Psi \ ,  &\partial_\mu J^\mu_A = &2im\mybar\Psi\gamma_5\Psi\ .&&
\eqn{noether}\end{align}

Some comments are in order:
\begin{itemize}
\item Eqn.~(\ref{eq:noether}) is not the whole story!    We will soon talk about additional contributions to the divergence of the axial current from the regulator, called anomalies, which do not decouple as the regulator is removed.
\item The fact that the fermion mass breaks chiral symmetry means that fermion masses get multiplicatively renormalized, which means that 
fermions can naturally be light (unlike scalars in most theories); more on this later.
\item The variation of a general fermion bilinear $\mybar \Psi X \Psi$ under chiral symmetry is
\beq
\delta \mybar \Psi X \Psi = i \mybar\Psi\{ \gamma_5,X\}\Psi\ .
\eeq
This will vanish if $X$  can be written as  the product of an odd number of gamma matrices.  In any even dimension the  chirally invariant bilinears include currents, with  $X=\gamma^\mu$ or $X=\gamma^\mu\Gamma$, while the bilinears which transform nontrivially under the chiral symmetry include mass terms,  $X= {\mathbf 1}, \Gamma$.  Thus gauge interactions can be invariant under chiral symmetry, while fermion masses are always chiral symmetry violating.  In $(3+1)$ dimensions, anomalous electromagnetic moment operators corresponding to $X=\sigma_{\mu\nu}, \sigma_{\mu\nu}\gamma_5$ are also chiral symmetry violating.
\item A more general expression for the classical divergence of the axial current for a bilinear action $\int \mybar \Psi D\Psi$ in any even dimension is
\beq
\partial_\mu J^\mu_A = i\mybar\Psi \{\Gamma, D\}\Psi\ .
\eqn{nonchiralD}
\eeq
On the lattice one encounters versions of the fermion operator $D$
which violate chiral symmetry even for a massless fermion.

\end{itemize}

The Lagrangian for $N_f$  flavors of massive Dirac fermions in odd $d$, coupled to some background gauge field may be written as
\beq
\CL =\left( \mybar\Psi^a_L i\Dslash \Psi^a_L +  \mybar\Psi^a_R i\Dslash \Psi^a_R\right) - \left(\mybar\Psi^a_L M_{ab}\Psi^b_R +\mybar\Psi^a_R M^\dagger_{ab}\Psi^b_L\right)\ .
\eqn{dl}\eeq
The index on $\Psi$ denotes flavor,  with $a,b = 1,\ldots N_f$, and $M_{ab}$ is a general complex mass matrix (no distinction between upper and lower flavor indices).
Again assuming the fermions to be in a complex representation of the gauge group, this theory is invariant under independent chiral transformations if the mass matrix vanishes:
\beq
\Psi_R^a \to U_{ab} \Psi^b_R\ ,\quad
\Psi^a_L \to V_{ab} \Psi^b_L\ ,\quad U^\dagger U = V^\dagger V = {\mathbf 1}\ .
\eqn{trans1}\eeq
where $U$ and $V$ are independent $U(N_f)$ matrices.  Since $U(N_f)= SU(N_f)\times U(1)$, it is convenient to write
\beq
U = e^{i(\alpha+\beta)} R\ ,\quad V=e^{i(\alpha-\beta)}L\ ,\quad R^\dagger R = L^\dagger L =  {\mathbf 1}\ ,\quad |R| = |L|=1\ ,
\eqn{trans2}\eeq
so that the symmetry group is $SU(N_f)_L\times SU(N_f)_R\times U(1)\times U(1)_A$ with $L\in SU(N_f)_L$, $R\in SU(N_f)_R$.

If we turn on the mass matrix, the chiral symmetry is explicitly broken, since the mass matrix couples left- and right-handed fermions to each other.  If $M_{ab} = m \delta_{ab}$ then the ``diagonal'' or ``vector''  symmetry $SU(N_f)\times U(1)$ remains unbroken, where $SU(N_f)\subset SU(N_f)_L\times SU(N_f)_R$ corresponding to the transformation \eq{trans1}, \eq{trans2} with $L=R$.  If $M_{ab}$ is diagonal but with unequal eigenvalues, the symmetry may be broken down as far as $U(1)^{N_f}$, corresponding to independent phase transformations of the individual flavors.  With additional flavor-dependent interactions, these symmetries may be broken as well.

\section{Weyl fermions}
\label{cgt}
\subsection{Lorentz group as $SU(2)\times SU(2)$ and Weyl fermions}
We have seen that Dirac fermions in even dimensions form a reducible representation of the Lorentz group.  Dirac notation is convenient when both LH and RH parts of the Dirac spinor transform as the same complex representation under a gauge group, and when there is a conserved fermion number.  This sounds restrictive, but applies to QED and QCD.  For other applications --- such as chiral gauge theories (where LH and RH fermions carry different gauge charges, as under $SU(2)\times U(1)$),  or when fermion number is violated (as is the case for neutrinos with a Majorana mass), or when fermions transform as a real representation of  gauge group --- then it is much more convenient to use irreducible fermion representations, called Weyl fermions.  

The six generators of the Lorentz group may be chosen to be the three Hermitian generators of rotations $J_i$, and the three anti-Hermitian generators of boosts $K_i$, so that an arbitrary Lorentz transformation takes the form
\beq
\Lambda = e^{i(\theta_i J_i + \omega_i K_i)}\ .
\eeq
In terms of the $M_{\mu\nu}$ generators in \S\ref{Lorentz},
\beq
J_i = \half\epsilon_{0i\mu\nu}M^{\mu\nu}\ ,\qquad
K_i = M^{0i}\ .
\eeq
These generators have the commutation relations
\beq
[J_i,J_j]=i\epsilon_{ijk} J_k\ ,\quad
[J_i,K_j]=i\epsilon_{ijk} K_k\ ,\quad
[K_i,K_j]=-i\epsilon_{ijk} J_k\ .
\eeq
It is convenient to define different linear combinations of generators
\beq
A_i = \frac{J_i+i K_i}{2}\ ,\qquad B_i = \frac{J_i-i K_i}{2}\  ,
\eqn{ab}\eeq 
satisfying an  algebra that looks like $SU(2)\times SU(2)$, except for the fact that the generators \eq{ab} are not Hermitian and therefore the group is noncompact:
\beq
[A_i,A_j] = i \epsilon_{ijk}A_k\ , \quad [B_i,B_j] = i \epsilon_{ijk}B_k , \quad  [A_i,B_j] =0 \ .
\eeq 
Thus Lorentz representations may be labelled with two $SU(2)$ spins, $j_{A,B}$ corresponding to the two $SU(2)$s:  $(j_A,\,j_B)$, transforming as 
\beq
\Lambda(\vec\theta,\,\vec\omega) = D^{j_A}( \vec\theta -i\vec\omega) \times D^{j_B}( \vec\theta +\vec\omega)
\eqn{ltrans}\eeq
where the $D^{j}$ is the usual $SU(2)$ rotation in the spin $j$ representation;  boosts appear as imaginary parts to the rotation angle;  the $D^{j_A}$ and $D^{j_B}$ matrices act in different spaces and therefore commute.  For example, under a general Lorentz transformation, a LH Weyl fermion $\psi$ and a RH Weyl fermion $\chi$   transform as $\psi\to L\psi$, $\chi\to R\chi$, where
\beq
 L=e^{i(\vec\theta -i \vec\omega)\cdot\vec\sigma/2}\ ,\qquad R=e^{i(\vec\theta +i \vec\omega)\cdot\vec\sigma/2}\ .
\eqn{ldef}\eeq
Evidently the two types of fermions transform the same way under rotations, but differently under boosts.

The dimension of the $(j_A,\,j_B)$ representation is $(2j_A+1)(2j_B+1)$.  In this notation, the smaller irreducible Lorentz representations are labelled as:
\begin{align}
(0,0): &\quad\text{scalar}\cr
(\half,0)\ ,\, (0,\half): &\quad\text{LH and RH Weyl fermions}\cr
(\half,\half): &\quad\text{four-vector}\cr
(1,0)\ ,\, (0,1): &\quad\text{self-dual and anti-self-dual antisymmetric tensors}\nonumber
\end{align}
A Dirac fermion is the  reducible representation $(\half,0)\oplus (0,\half)$ consisting of a LH and a RH Weyl fermion.

Parity interchanges the two $SU(2)$s,  transforming a $(j_1,\,j_2)$ representation into  $(j_2,\,j_1)$.  Similarly, charge conjugation effectively flips the sign of $K_i$ in \eq{ltrans} due to the factor of $i$ implying that if a field $\phi$ transforms as $(j_1,\,j_2)$, then $\phi^\dagger$ transforms as $(j_2,j_1)$\footnote{For this reason, the combined symmetry CP does not alter the particle content of a chiral theory, so that CP violation muast arise from complex coupling constants.}. Therefore a theory of $N_L$ flavors of LH Weyl fermions $\psi_i$,  and $N_R$ flavors of RH Weyl fermions $\chi_a$ may be recast as a theory of $(N_L+N_R)$ LH fermions by defining $\chi_a\equiv \omega_a^\dagger$.  The fermion content of the theory can be described entirely in therms of LH Weyl fermions then, $\{\psi_i, \omega_a\}$;  this often simplifies the discussion of parity violating theories, such as the Standard Model or Grand Unified Theories.  Note that if the RH $\chi_a$ transformed under a gauge group as representation $R$, the conjugate fermions $\omega_a$ transform under the conjugate representation $\mybar R$.

For example, QCD written in  terms of Dirac fermions has the Lagrangian:
\beq
\CL = \sum_{i=u,d,s\ldots} \mybar\Psi_n( i\Dslash -m_n)\Psi_n \ ,
\eeq
where $D_\mu$ is the $SU(3)_c$ covariant derivative, and the $\Psi_n$ fields (both LH and RH components) transform as a $3$ of $SU(3)_c$.  However, we could just as well write the theory in terms of the LH quark fields $\psi_n$ and the LH anti-quark fields $\chi_n$.  Using the $\gamma$-matrix basis in \eq{d4chiral}, we write the Dirac spinor $\Psi$ in terms of two-component LH spinors $\psi$ and $\chi$ as
\beq
\Psi = \begin{pmatrix} -\sigma_2 \chi^\dagger\cr \psi
\end{pmatrix}\ .
\eeq
Note that $\psi$ transforms as a $3$ of $SU(3)_c$, while $\chi$ transforms as a $\mybar 3$.
Then the kinetic operator becomes (up to a total derivative)
\beq
\mybar\Psi i\Dslash\Psi = \psi^\dagger i D_\mu \sigma^\mu \psi +  \chi^\dagger i D_\mu \sigma^\mu \chi\ ,\qquad \sigma^\mu \equiv\{{\mathbf 1},-\vec\sigma\}\ ,
\eqn{smdef}\eeq
and the mass terms become
\beq
\mybar\Psi_R\Psi_L  &=& \chi \sigma_2 \psi=  \psi\sigma_2 \chi\cr
\mybar\Psi_L\Psi_R  &=& \psi^\dagger \sigma_2 \chi^\dagger=  \chi^\dagger \sigma_2 \psi^\dagger \ ,
\eeq
where I used the fact that fermion fields anti-commute.
Thus a Dirac mass in terms of Weyl fermions is just
\beq
m\mybar\Psi \Psi = m ( \psi\sigma_2 \chi + h.c)\ ,
\eeq
and preserves a fermion number symmetry where $\psi$ has charge $+1$ and $\chi$ has charge $-1$.  On the other hand, one can also write down a Lorentz invariant mass term of the form
\beq
m (\psi\sigma_2\psi + h.c.)
\eeq
which violates fermion number by two units;  this is a Majorana mass, which is clumsy to write in Dirac notation.  Experimentalists are trying to find out which form neutrino masses have --- Dirac, or Majorana?  If the latter, lepton number is violated by two units and could show up in neutrinoless double beta decay, where a nucleus decays by emitting two electrons and no anti-neutrinos.

The Standard Model is a relevant example of a chiral gauge theory.  Written in terms of LH Weyl fermions, the quantum numbers of a single family under $SU(3)\times SU(2)\times U(1)$ are:
\begin{align}
Q &= (3,2)_{+\frac{1}{6}} \qquad & L&=(1,2)_{-\frac{1}{2}}\cr
U^c&=(\mybar 3,1)_{-\frac{2}{3}} & E^c&=(1,1)_{+1}\cr
D^c &= (\mybar 3,1)_{+\frac{1}{3}} \ .&
\eqn{SM}\end{align}
Evidently this is a complex representation and chiral. If neutrino masses are found to be Dirac in nature (i.e. lepton number preserving) then a partner for the neutrino must be added to the theory, the ``right handed neutrino", which can be described by a LH Weyl fermion which is neutral under all Standard Model gauge interactions,  $N=(1,1)_0$.
 
  If  unfamiliar with two-component notation, you can find all the details  in Appendix~A of Wess and Bagger's classic book on supersymmetry \cite{Wess:1992cp};  the notation used here differs  slightly as I use the metric and $\gamma$-matrix conventions of Itzykson and Zuber \cite{Itzykson:1980rh}, and write out the $\sigma_2$ matrices explicitly.
  
\begin{exercisebn}
\label{ex2b}
Consider a theory of $N_f$ flavors of Dirac fermions in a real or pseudo-real representation of some gauge group.  (Real representations combine symmetrically to form an invariant, such as a triplet of $SU(2)$; pseudo-real representations combine anti-symmetrically, such as a doublet of $SU(2)$). Show that if the fermions are massless the action exhibits a $U(2N_f)= U(1)\times SU(2N_f)$ flavor symmetry at the classical level (the $U(1)$ subgroup being anomalous in the quantum theory). If the fermions condense as in QCD, what is the symmetry breaking pattern? How do the resultant Goldstone bosons transform under the unbroken subgroup of $SU(2N_f)$?
 \end{exercisebn}
  \begin{exercisenn}
To see how the $(\half,\half)$ representation behaves like a four-vector, consider the $2\times 2$ matrix $P = P_\mu \sigma^\mu$, where $\sigma^\mu$ is given in \eq{smdef}.  Show that the transformation $P\rightarrow L P L^\dagger$ (with  $\det L=1$) preserves the Lorentz invariant inner product $P_\mu P^\mu= (P_0^2 - P_i P_i)$.   Show that with $L$ given by \eq{ldef},  $P_\mu$ transforms properly like a four-vector.
  \end{exercisenn}
  \begin{exercisenb}
Is it possible  to write down an anomalous electric or magnetic moment operator in a theory of a single charge-neutral Weyl fermion?  
  \end{exercisenb}

\section{Chiral symmetry and mass renormalization}
\label{mass}
Some operators in a Lagrangian suffer from additive renormalizations, such as the unit operator (cosmological constant) and scalar mass terms, such as the Higgs mass in the Standard Model,  $|H|^2$.  Therefore, the mass scales associated with such operators will naturally be somewhere near the UV cutoff of the theory, unless the bare  couplings of the theory are fine-tuned to cancel radiative corrections.  Such fine tuning problems have obsessed particle theorists since the work of Wilson and 't Hooft on renormalization and naturalness in the 1970s. However, such intemperate behavior will not occur for operators which violate a symmetry respected by the rest of the theory:  if the bare couplings for such operators were set to zero, the symmetry would ensure they could not be generated radiatively in perturbation theory. Fermion mass operators generally fall into this benign category. 

\begin{figure}[t]
\begin{center}
  \includegraphics[width=0.40\textwidth]{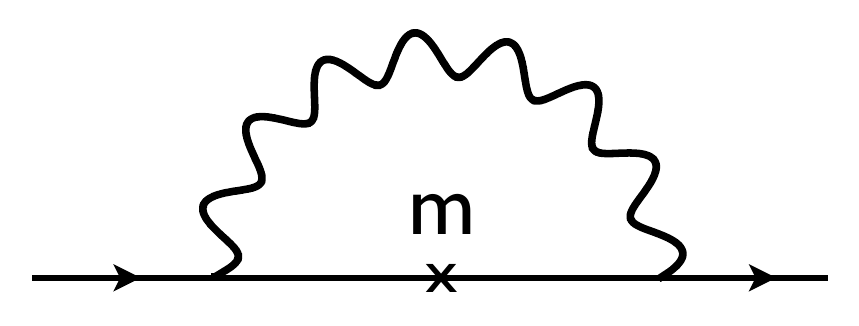}
  \end{center}
\caption{\it One-loop renormalization of the electron mass in QED due to photon exchange.  A mass operator flips chirality, while gauge interactions do not.  A contribution to the electron mass requires an odd number of chirality flips, and so there has to be at least one insertion of the electron mass in the diagram:  the electron mass is multiplicatively renormalized.  A  scalar interaction flips chirality when the scalar is emitted, and flips it back when the scalar is absorbed, so replacing the photon with a scalar in the above graph again requires a fermion mass insertion to contribute to mass renormalization.}
\label{fig:massrenorm}
\end{figure}

Consider the following toy model:  QED with a charge-neutral complex scalar field coupled to the electron:
\beq
\CL \mybar\Psi (i\Dslash-m)\Psi + |\partial\phi\vert^2 - \mu^2|\phi|^2 -g|\phi|^4 + y\left(\mybar\Psi_R \phi \Psi_L + \mybar\Psi_L\phi^*\Psi_R\right)\ .
\eeq
Note that in the limit $m\to 0$ this Lagrangian respects a chiral symmetry $\Psi\to e^{i\alpha\gamma_5}\Psi$, $\phi\to e^{-2i\alpha}\phi$.  The symmetry  ensures that if $m=0$, a mass term for the fermion would not be generated radiatively in perturbation theory.  With $m\ne 0$, this means that any renormalization of $m$ must be proportional to $m$ itself (i.e. $m$ is ``multiplicatively renormalized").  This is evident if one traces chirality through the Feynman diagrams; see Fig.~\ref{fig:massrenorm}.  Multiplicative renormalization implies that the fermion mass can at most depend logarithmically on the cutoff (by dimensional analysis):  $\delta m \sim (\alpha/4\pi) m\ln m/\Lambda$.

\begin{figure}[t]
\begin{center}
  \includegraphics[width=0.40\textwidth]{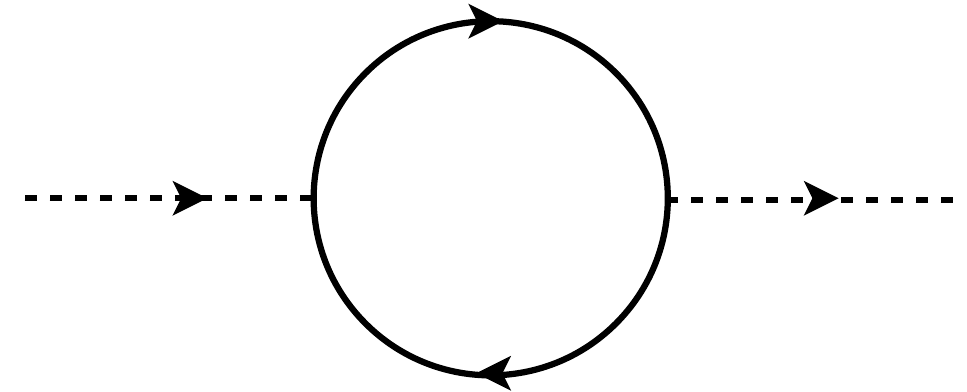}
  \end{center}
\caption{\it One-loop additive renormalization of the scalar mass due to a quadratically divergent fermion loop. }
\label{fig:massrenorm2}
\end{figure}

In contrast, the scalar mass operator $|\phi|^2$ does not violate any symmetry and therefore suffers from additive renormalizations, such as through the graph in  Fig.~\ref{fig:massrenorm2}.  By dimensional analysis, the scalar mass operator can have a coefficient that scales quadratically with the cutoff:  $\delta\mu^2 \sim (y^2/16\pi^2)  \Lambda^2$.  This is called an additive renormalization, since $\delta\mu^2$ is not proportional to $\mu^2$.  It is only possible in general to have a scalar in the spectrum of this theory with mass much lighter than $y\Lambda/4\pi$ if the bare couplings are finely tuned to cause  large radiative corrections to cancel.  When referring to the Higgs mass in the Standard Model, this is called the hierarchy problem.

If chiral symmetry is broken by operators other than the mass term, then the fermion mass will no longer in general be multiplicatively renormalized, and fine tuning may be necessary.  This is particularly true if chiral symmetry is broken by ``irrelevant" operators.  Consider adding to QED a dimension five operator of the form
$
W=\frac{r}{\Lambda}\mybar\Psi D_\mu D^\mu \Psi \ ,
$
where $r$ is a dimensionless coupling. This operator breaks chirality and therefore one can substitute it for the mass operator in Fig.~\ref{fig:massrenorm}; 
an estimate of the diagram then gives an additive renormalization of the fermion mass, $\delta m \sim (\alpha/4\pi) r \Lambda$.  Thus unless $r$ is extremely small ({\it e.g.} $r\lesssim m/\Lambda$) the chiral symmetry breaking effects of this operator will be important and fine tuning will be necessary to ensure a light fermion in the spectrum.  This example is relevant to Wilson's method for putting fermions on the lattice, which does not respect chiral symmetry and entails adding to the action a lattice version of $W$, where $a\sim 1/\Lambda$ is the lattice spacing and $r\sim 1$. Therefore Wilson fermions acquire an $O(1/a)$ correction to their mass which needs to be canceled by a bare contribution in order to describe a world with light fermions.

\section{Chiral symmetry in QCD}
\label{opmix}

\subsection{Chiral symmetry breaking and Goldstone bosons}
\label{chisb}
So far the discussion of chiral symmetry in terms of the effect on fermion masses has been appropriate for a weakly coupled theory.  As was presented in M. Golterman's lectures, the low energy spectrum of QCD is described by a chiral Lagrangian, encoding the interactions of the meson octet which are the approximate Nambu-Goldstone bosons of spontaneously broken $SU(3)_L\times SU(3)_R$ chiral symmetry \footnote{With six favors of quarks in QCD, one might ask why an $SU(3)$ chiral symmetry instead of $SU(2)$ or $SU(6)$. The point is that the chiral symmetry is broken by quark masses, and whether the breaking is large or small depends on the ratio $m_q/\Lambda_{QCD}$, where here $\Lambda_{QCD}$ is some strong interaction scale in the 100s of MeV.  The $u$ and $d$ quarks are much lighter than $\Lambda_{QCD}$, the strange quark is borderline, and the $c,b,t$ quarks are much heavier.  Therefore $SU(2)\times SU(2)$ is a very good symmetry of QCD; $SU(3)\times SU(3)$ is a pretty good symmetry of QCD, but assuming chiral symmetry for the heavier quarks is not justified.  Radiative corrections in the baryon sector go as $\sqrt{m_q/\Lambda_{QCD}}$ and so there even $SU(3)\times SU(3)$ does not appear to be very reliable.}.  The Goldstone bosons would be massless in the limit of exact chiral symmetry, and so tuning away the leading  finite lattice space correction for Wilson fermions can be accomplished by tuning the bare quark mass to eliminate the $1/a$ dependence of the square of the pion mass.  

  In contrast to QCD, $N=1$ super Yang Mills theory has a single Weyl fermion (the gaugino) transforming as an adjoint under the gauge group; the theory has a $U(1)_A$ symmetry at the classical level  --- phase rotations of the gaugino --- but it is broken by anomalies to a discrete symmetry.  This discrete symmetry is then spontaneously broken by a gluino condensate, but without any continuous symmetries, no Goldstone bosons are produced.  What should the spectrum of this theory look like?  Presumably a bunch of massive boson and fermion glueball-like states.  They will form degenerate supersymmetric multiplets when the gluino mass is tuned to zero,   but there is no particle that becomes massless in the chiral limit in this case, and therefore  tuning the bare mass is difficult.

After tuning away the $O(1/a)$ mass correction,   there remain for non-chiral lattice fermions the dimension-5 chiral symmetry violating operators in the Symanzik action which require $O(a)$ tuning, as discussed by Golterman.  In contrast, chiral fermions receive finite lattice corrections only at $O(a^2)$, simply because one cannot write down a dimension-5 chiral symmetry preserving operator in QCD.

\subsection{Operator mixing}
\label{opmix}

One encounters additional factors of $1/a$ when computing weak processes.  One of the most curious feature of the strong interactions is the $\Delta I=1/2$ rule, which is the observation that $\Delta s=1$ transitions in nature are greatly enhanced when they change isospin by $\Delta I=1/2$, in comparison to $\Delta I=3/2$.   For example, one requires for the amplitudes for kaon decay $K\to \pi\pi$:
\beq
\frac{\CA(\Delta I=1/2)}{\CA(\Delta I=3/2)}\simeq 20\ .
\eqn{I12}
\eeq
To compute this in the standard model, one starts with four-quark operators generated by $W$-exchange, which can be written as the linear combination of two operators
\beq
\CL_{\Delta S=1}&= &- V_{ud}V^*_{us} \frac{G_F}{\sqrt{2}}\left[C_+(\mu, M_w)\CO^+ + C_-(\mu, M_w)\CO^-\right]\ ,\cr
\CO^{\pm} &=& \left[(\mybar sd)_L(\mybar u u)_L \pm (\mybar su)_L(\mybar u d)_L\right] - \left[u\leftrightarrow c\right]\ ,
\eqn{dscont}
\eeq
where $(\mybar q q')_L \equiv  (\bar q \gamma^\mu P_L q')$.  If one ignores the charm quark contribution, the $\CO^-$ transforms as an 8 under $SU(3)_f$, while $\CO^+$ transforms as a 27; therefore  $\CO^{-} $ is pure $I=1/2$, while $\CO^+ $ is a mix of $I=3/2$ and $I=1/2$.  The full $\CO^{\pm} $ operators are in $SU(4)_f$ multiplets;  while $SU(4)_f$ is not a good symmetry of the spectrum, it is only broken by quark masses which do not effect the log divergences of the theory.  Thus the running of the operators respect $SU(4)_f$ down to $\mu=m_c$, and there is no mixing between $\CO^{\pm} $. At the weak scale $\mu=M_W$, one finds $|C^+/C^-|=1 + O(\alpha_s(M_W)$, showing that there is no $\Delta I=1/2$ enhancement intrinsic to the weak interactions.  One then scales these operators down to $\mu\sim 2$ GeV in order to match onto the lattice theory; using the renormalization group to sum up leading $\alpha_s \ln \mu/M_W$ corrections gives an enhancement $|C^+/C^-|\simeq 2$ --- which is in the right direction, but not enough to explain  \eq{I12}, which should then either be coming from QCD at long distances, or else new physics!  This is a great problem for the lattice to resolve.

A wonderful feature about using dimensional regularization and $\mybar{MS}$ in the continuum is that an operator will never mix with another operator of lower dimension.  This is because there is no UV mass scale in the scheme which can make up for the miss-match in operator dimension.  This is not true on the lattice, where powers of the inverse lattice spacing $1/a$ can appear.  In particular, the the dimension-6 four fermion operators $\CO^\pm $ could in principle mix with dimension-3 two fermion operators.  The only  $\Delta S=1$ dimension-3 operator that could arise is $\mybar s \gamma_5 d$, which is also $\Delta I=\half$ \footnote{The operator  $\mybar s d$ is removed by rediagonalizing the quark mass matrix and does not give rise to $K\pi\pi$.}.  If the quarks were massless, the lattice theory would possess an exact  discrete ``SCP" symmetry under which on interchanges $s\leftrightarrow d$ and performs a CP transformation to change LH quarks into LH anti-quarks; the operators $\CO_\pm$ are even under SCP while $\mybar s \gamma_5 d$ is odd, to the operator that could mix on the lattice is 
\beq
\CO_p=(m_s-m_d)\,\mybar s\gamma_5 d\ .
\eqn{d3op}
\eeq
 In a theory where the quark masses are the only source of chiral symmetry breaking, then $\CO_p=\partial_\mu A_\mu^{\mybar s d}$, the divergence of the $\Delta S=1$ axial current.  Therefore on-shell matrix elements of this operator vanish, since the derivative gives $(p_K -p_{2\pi})=0$, i.e. no momentum is being injected by the weak interaction.  We can ignore $\CO_p$ then when the $K\to\pi\pi$ amplitude is measured with chiral lattice fermions with on-shell momenta.  
 
 For a lattice theory without chiral symmetry, $\CO_p = \partial_\mu A_\mu^{\mybar s d} + O(a)$ and so has a  nonvanishing $O(a)$ matrix element.  In this case operators $\CO_\pm$ from \eq{dscont} in the continuum match onto the lattice operators
\beq
\CO^\pm(\mu) = Z^\pm(\mu a, g_0^2) \left[ \CO^\pm(a) + \frac{C_p^\pm}{a^2} \CO_p\right] + \CO(a)\ .
\eeq
In general then one would need to determine the coefficient $C_p^\pm$ to $O(a)$ in order to determine the $\Delta I=\half$ amplitude for $K\to \pi\pi$ to leading order in an $a$-expansion, which is not really feasible.  Other weak matrix elements such as $B_K$ and $\epsilon'/\epsilon$ similarly benefit from the use of lattice fermions with good chiral symmetry.

 \section{Fermion determinants in Euclidian space}

Lattice computations employ Monte Carlo integration, which requires a positive integrand that can be interpreted as a probability distribution.  While not sufficient for a lattice action to yield a positive measure, it is certainly necessary  for the continuum theory one is approximating to have this property.  Luckily, the fermion determinant for vector-like gauge theories (such as QCD), $\det (\Dslash + m)$,  has this property in Euclidian space.    Since $\Dslash^\dagger = -\Dslash$ and  $\{\Gamma,\Dslash\}=0$, it follows that there exist eigenstates $\psi_n$ of $\Dslash$ such that
\beq
\Dslash \psi_n = i\lambda_n \psi_n\ ,\qquad \Dslash \Gamma\psi_n = -i\lambda_n \Gamma\psi_n\ ,\qquad \lambda_n\text{ real.}
\eeq
For nonzero $\lambda$,  $\psi_m$ and $\Gamma\psi_n$ are all mutually orthogonal and we see that the eigenvalue spectrum contains $\pm i\lambda_n$ pairs.  On the other hand, if $\lambda_n=0$ then $\psi_n$ can be an eigenstate of $\Gamma$ as well, and $\Gamma\psi_n$ is not an independent mode.  Therefore
\beq
\det (\Dslash + m)= \prod_{\lambda_n>0} (\lambda_n^2+m^2)\times \prod_{\lambda_n=0} m 
\eeq
which is real and for positive $m$ is positive for all gauge fields.

What about a chiral gauge theory?  The fermion Lagrangian for a LH Weyl fermion in Euclidian space looks like $\mybar\psi D_L\psi$ with $D_L = D_\mu \sigma_\mu$ and (in the chiral basis \eq{d4chiral}, continued to Euclidian space) $\sigma_\mu =\{1,i\vec\sigma\} $.   Note that $D_L$ has no nice hermiticity properties, which means its determinant will be complex,  its right eigenvectors and left eigenvectors will be different,  and its eigenvectors will not be mutually orthogonal.  Furthermore, $D_L$ is an operator which maps vectors from the space $\CL$ of LH Weyl fermions to the space $\CR$ of RH Weyl fermions.  In Euclidian space, these spaces are unrelated and transform independently under the $SU(2)\times SU(2)$ Lorentz transformations.  Suppose we have an orthonormal basis $\ket{n,\CR}$ for the RH Hilbert space and $\ket{n,\CL}$ for the LH Hilbert space;  we can expand our fermion integration variables as
\beq
\psi = \sum_n c_n \ket{n,\CL}\ ,\qquad \mybar\psi = \sum_n \mybar c_n \bra{n,\CR}
\eeq
so that
\beq
\int [d\psi][d\mybar\psi] \, e^{-\int\mybar\psi D_L \psi} = \det_{mn} \expect{m,\CR}{D_L}{n,\CL}\ .
\eeq
However, the answer we get will depend on the basis we choose.  For example, we could have chosen a different orthonormal basis for the $\CL$ space $\ket{n',\CL} = \CU_{n'n}\ket{n,\CL}$ which differed from the first by a unitary transformation $\CU$;  the resultant determinant would differ by a factor $\det\CU$, which is a phase.  If this phase were a number, it would not be an issue --- but it can in general be a functional of the background gauge field, so that different choices of phase for $\det D_L$ lead to completely different theories.


\begin{figure}[t]
\begin{center}
  \includegraphics[width=0.70\textwidth]{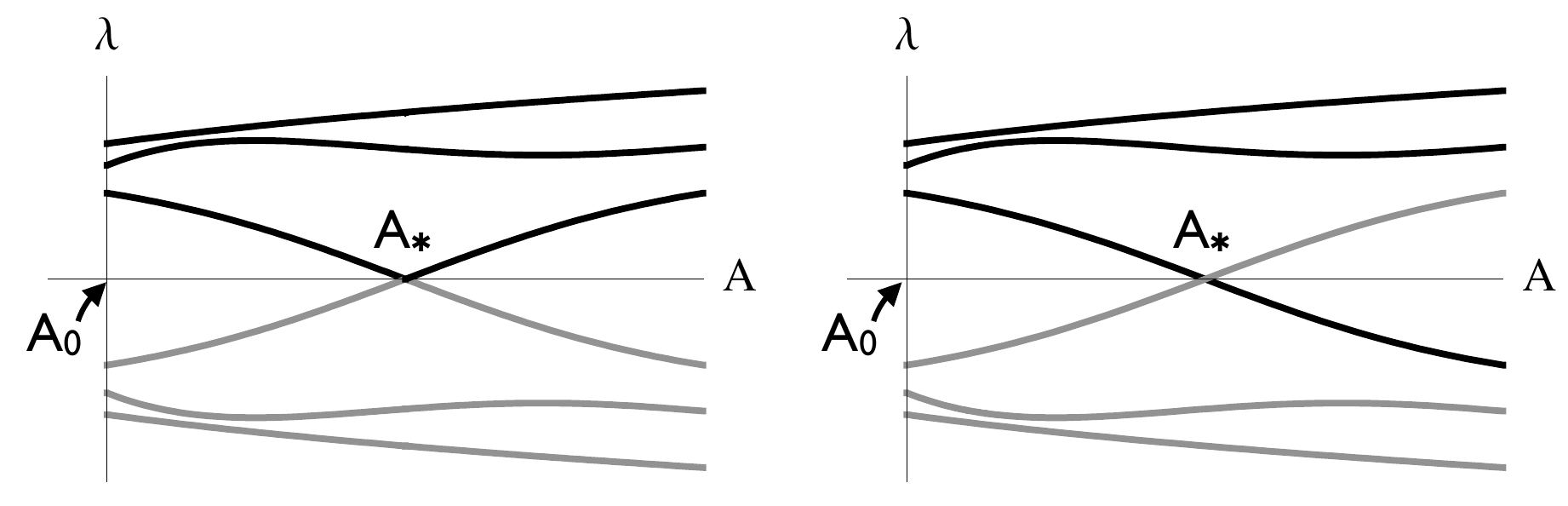}
  \end{center}
\caption{\it The eigenvalue flow of the Dirac operator as a function of gauge fields, and two unsatisfactory ways to define the Weyl fermion determinant $\det D_L$ as a square root of $\det \Dslash$.  The expression $\sqrt{\left|\det\Dslash\right|}$ corresponds to the picture on the left, where $\det D_L$ is defined as the product of positive eigenvalues of $\Dslash$;  this definition is nonanalytic at $A_*$. The picture on the right  corresponds to the product of half the eigenvalues, following those which were positive at some reference gauge field $A_0$.  This definition is analytic, but not necessarily local.  Both definitions are gauge invariant, which is incorrect for an anomalous fermion representation.}
\label{fig:eigenvalues}
\end{figure}


We do know that if $D_R$ is the fermion operator for RH Weyl fermions in the same gauge representation as $D_L$, then $\det D_R = det D_L^*$ and that $\det D_R\det D_L = \det \Dslash$.  Therefore the the norm of $\left\vert \det D_L\right\vert$ can be defined as
\beq
\det D_L = \sqrt{\left|\det\Dslash\right|} e^{iW[A]}
\eeq
where the phase $W[A]$ is a functional of the gauge fields.  What do we know about $W[A]$?
\begin{enumerate}
\item Since $\det\Dslash$ is gauge invariant, $W[A]$ should be gauge invariant unless the fermion representation has a gauge anomaly, in which case it should correctly reproduce that anomaly;
\item It should be analytic in the gauge fields, so that the computation of gauge field correlators (or the gauge current) are well defined.
\item It should be a local functional of the gauge fields.
\end{enumerate}

In Fig.~\ref{fig:eigenvalues} I show two possible ways to define $\det D_L$, neither of which satisfy the above criteria.  The naive choice of just setting $W[A]=0$ not only fails to reproduce the anomaly (if the fermion representation is anomalous) but is also nonanalytic and nonlocal.  It corresponds to taking the product of all the positive eigenvalues $\lambda_n$ of $\Dslash$ (up to an uninteresting overall  constant phase).  This definition is seen to be nonanalytic where eigenvalues cross zero.  Another definition might be to take the product of positive eigenvalues at some reference gauge field $A_0$, following those eigenvalues as they cross zero;  this definition is analytic, but presumably not local, and is always gauge invariant.

There has been quite a few papers on how to proceed in defining this phase $W[A]$ in the context of domain wall fermions, including a rather complicated explicit construction for $U(1)$ chiral gauge theories on the lattice \cite{Luscher:1998du,Luscher:2000hn};  however, even if a satisfactory definition of $W[A]$ is devised, it could be impossible to simulate using Monte Carlo algorithms due to the complexity of the fermion determinant.

\section{Parity and fermion mass in odd dimensions}
\label{parity}

In these lectures I will be discussing  fermions in $(2k+1)$ dimensions with a spatially varying mass term which vanishes in some $2k$-dimensional region; in such cases we find chiral modes of a $2k$-dimensional effective theory bound to this mass defect.  Such an example could arise dynamically when fermions have a Yukawa coupling to a real scalar $\phi$ which spontaneously breaks a discrete symmetry, where the surface with  $\phi=0$ forms a domain wall between two different phases; for this reason such fermions are called domain wall fermions, even though we will be putting the spatially dependent mass in by hand and not through spontaneous symmetry breaking.

To study domain wall fermions it is useful to say a few words about fermions in odd dimensions
where  there is no analogue of $\Gamma$ and therefore there is no such thing as chiral symmetry.  Nevertheless, fermion masses still break a symmetry:  parity.  In a theory with parity symmetry one has extended the Lorentz group  to include improper rotations: spatial rotations $R$ for which the determinant of $R$ is negative.   Parity can be defined as a transformation where  an odd number of the spatial coordinates flip sign.  In even dimensions parity can be the transformation $\bfx \to -\bfx$ and
\beq
 \Psi(\bfx,t) \to \gamma^0\Psi(-\bfx,t)\qquad (\text{parity, }d\text{ even})\ .
\eeq
Note that under this transformation $\Psi_L$ and $\Psi_R$ are exchanged and that a Dirac mass term is parity invariant.

However, in odd dimensions the transformation  $\bfx \to -\bfx$ is just a rotation;  instead we can define parity as the transformation which just flips  the sign of one coordinate $x^1$, and 
\beq
\Psi(\bfx,t) \to \gamma^1\Psi(\tilde\bfx,t)\ ,\qquad \tilde\bfx=(-x^1,x^2,\ldots,x^{2k})
\eeq
Remarkably, a Dirac mass term flips sign under parity in this case; and since there is no chiral symmetry in odd  $d$ to rotate the phase of the mass matrix, the sign of the quark mass is physical, and a parity invariant theory of massive quarks must have them come in pairs with masses $\pm M$, with parity interchanging the two.

\section{Fermion masses and regulators}
\label{regs}
We have seen that theories of fermions in any dimension can possess symmetries which forbid masses -- chiral symmetry in even dimensions and parity in odd dimensions.  This property obviously can have a dramatic impact on the spectrum of a theory.  Supersymmetry ingeniously puts fermions and bosons in the same supermultiplet, which allows scalars to also enjoy the benefits of chiral symmetry, which is one reason theorists have been so interested in having supersymmetry explain why the Higgs boson of the standard model manages to be so much lighter than the Planck scale.  However, precisely because mass terms violate these symmetries, it is difficult to maintain them in a regulated theory.  After all, one regulates a theory by introducing a high mass scale in order to eliminate UV degrees of freedom in the theory, and this mass scale will typically violate chiral symmetry in even dimensions, or parity in odd.  This gives rise to ``anomalous" violation of the classical fermion symmetries, my next topic.

%
%

\chapter{Anomalies}
\label{anomalies}

\section{The $U(1)_A$ anomaly in 1+1 dimensions}
\label{d2anomalies}

One of the fascinating features of chiral symmetry is that sometimes it is not a symmetry of the quantum field theory even when it is a symmetry of the Lagrangian.  In particular, Noether's theorem can be modified in a theory with an infinite number of degrees of freedom; the modification is called ``an anomaly".  Anomalies turn out to be very relevant both for phenomenology, and for the implementation of lattice field theory.  The reason anomalies affect chiral symmetries is that regularization requires a cut-off on the infinite number of modes above some mass scale, while chiral symmetry is incompatible with fermion masses\footnote{Dimensional regularization is not a loophole, since chiral symmetry cannot be analytically continued away from odd space dimensions.}. 

\begin{figure}[t]
\begin{center}
  \includegraphics[width=0.70\textwidth]{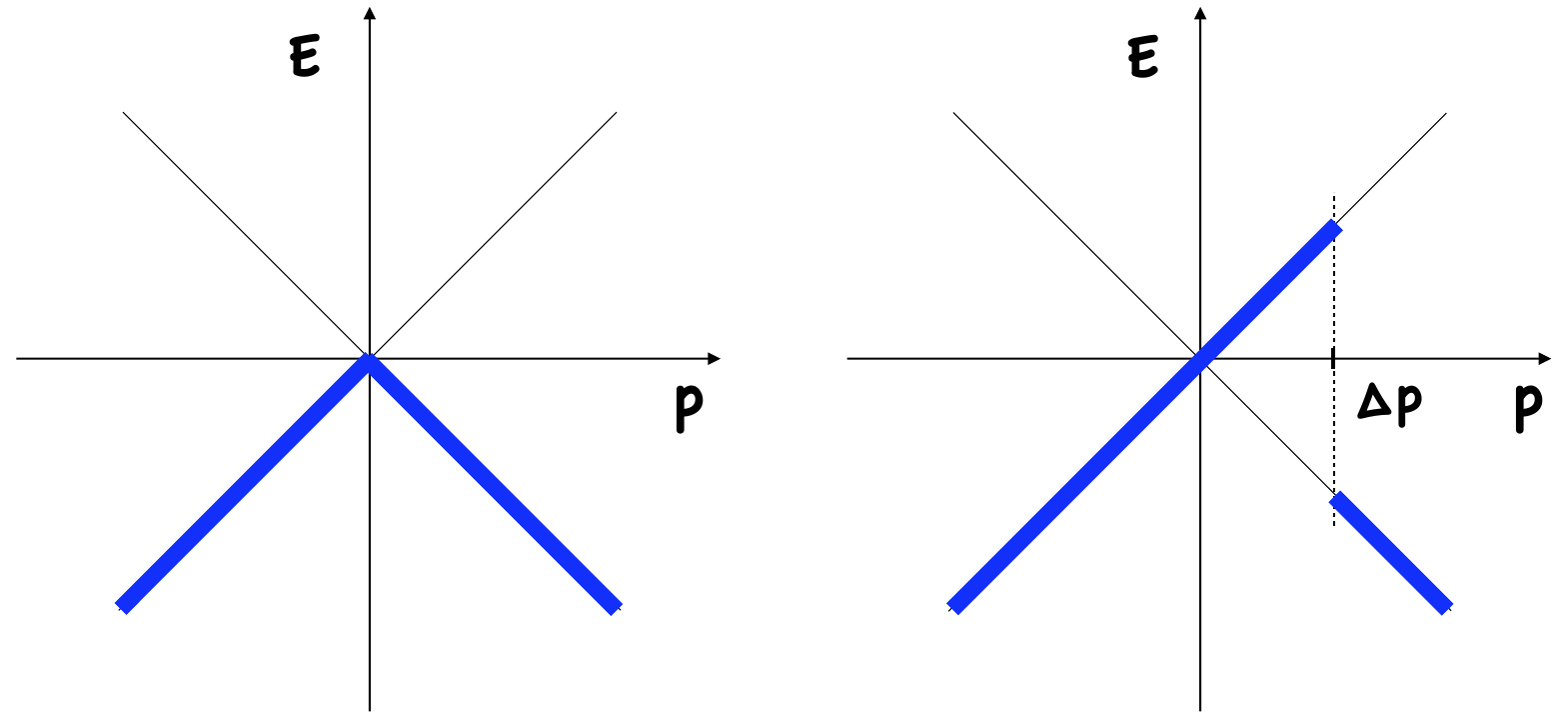}
  \end{center}
\caption{\it On the left: the ground state for a theory of a single massless Dirac fermion in $(1+1)$ dimensions;  on the right: the theory after application of an adiabatic electric field with all states shifted to the right by $\Delta p$, given in \eq{dp}. Filled states are indicated by the heavier blue lines. }
\label{fig:2Danomaly}
\end{figure}

Anomalies can be seen in many different ways.  I think the most physical is to look at what happens to the ground state of a theory with a single flavor of massless Dirac fermion in $(1+1)$ dimensions in the presence of an electric field.  Suppose one adiabatically turns on a constant positive electric field $E(t)$, then later turns it off;  the equation of motion for the fermion is \footnote{While in much of these lectures I will normalize gauge fields so that $D_\mu = \partial_\mu +i A_\mu$, in this section I need to put the gauge coupling back in.  If you want to return to the nicer normalization, set the gauge coupling to unity, and put a $1/g^2$ factor in front of the gauge action.}
$
\frac{d p}{dt} = e E(t)
$ and the total change in momentum is 
\beq
\Delta p = e\, \int E(t)\, dt\ .
\eqn{dp}\eeq
Thus the momenta of both left- and right-moving modes increase; if one starts in the ground state of the theory with filled Dirac sea, after the electric field has turned off, both the right-moving and left-moving sea has shifted to the right as in Fig.~\ref{fig:2Danomaly}. The the final state differs from the original by the creation of particle- antiparticle pairs:  right moving particles and left moving antiparticles.  Thus while there is a fermion  current in the final state, fermion number has not changed.  This is what one would expect from conservation of the $U(1)$ current:
\beq
\partial_\mu J^\mu =0\ ,
\eeq
However, recall that right-moving and left-moving particles have positive and negative chirality respectively; therefore the final state in Fig.~\ref{fig:2Danomaly} has net axial charge, even though the initial state did not.  This is peculiar, since the coupling of the electromagnetic field in the Lagrangian does not violate chirality.  We can quantify the effect: if we place the system in a box of size $L$ with periodic boundary conditions, momenta are quantized as $p_n= 2\pi n/L$.  The change in axial charge is then
\beq
\Delta Q_A = 2 \frac{\Delta p}{2\pi/L} = \frac{e}{\pi} \int d^2x \, E(t) = \frac{e}{2\pi} \int d^2x \, \epsilon_{\mu\nu} F^{\mu\nu}\ ,
\eeq
where I expressed the electric field in terms of the field strength $F$, where $F^{01}=-F^{10}=E$.   This can be converted into the local equation using $\Delta Q_A =  \int d^2x \,\partial_\mu J^\mu_A $, a modification of \eq{noether}:
\beq
\partial_\mu J^\mu_A =2im\mybar\Psi\Gamma\Psi + \frac{e}{2\pi}  \epsilon_{\mu\nu} F^{\mu\nu} \ ,
\eqn{dsanomaly2}\eeq
where in the above equation I have included the classical violation due to a mass term as well.
The second term is the axial anomaly in $1+1$ dimensions; it would vanish for a nonabelian gauge field, due to the trace over the gauge generator.  

So how did an electric field end up violating chiral charge?  Note that this analysis relied on the Dirac sea being infinitely deep.  If there had been a finite number of negative energy states, then they would have shifted to higher momentum, but there would have been no change in the axial charge.  With an infinite number of degrees of freedom, though, one can have a ``Hilbert Hotel":  the infinite hotel which can always accommodate another visitor, even when full, by moving each guest to the next room and thereby opening up a room for the newcomer.  This should tell you that it will not be straightforward to represent chiral symmetry on the lattice: a lattice field theory approximates  quantum field theory with a finite number of degrees of freedom --- the lattice is a big hotel, but quite conventional. In such a hotel there can be no anomaly.


\begin{figure}[t]
\begin{center}
  \includegraphics[width=0.35\textwidth]{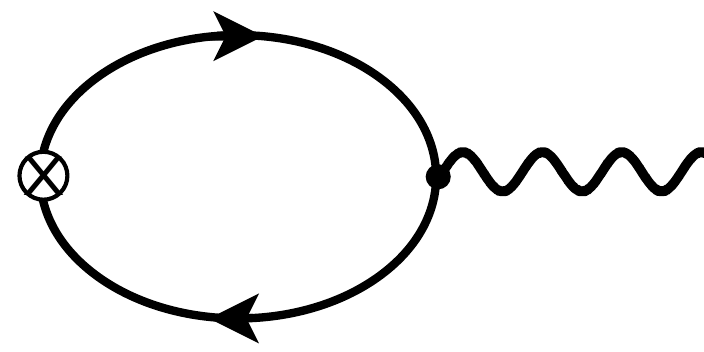}
  \end{center}
\caption{\it The anomaly diagram in 1+1 dimensions, with one Pauli-Villars loop and an insertion of  $2iM \mybar\Phi  \Gamma \Phi$ at the X. }
\label{fig:2Danomdiag}
\end{figure}

We can derive the anomaly in other ways, such as by computing the anomaly diagram Fig.~\ref{fig:2Danomdiag}, or by following Fujikawa \cite{Fujikawa:1979ay,Fujikawa:1980eg} and carefully accounting for the Jacobian from the measure of the path integral when performing a  chiral transformation. It is particularly instructive for our later discussion of lattice fermions to compute the anomaly in perturbation theory using Pauli-Villars regulators of mass $M$.  We replace our axial current by a regulated current
\beq
J^\mu_{A,\text{reg}} = \mybar\Psi \gamma^\mu \Gamma \Psi +  \mybar\Phi \gamma^\mu \Gamma \Phi\ ,
\eeq
where $\Phi$ is our Pauli-Villars field;  it follows then that
\beq
\partial_\mu    J^\mu_{A,\text{reg}}  = 2im  \mybar\Psi \Gamma \Psi + 2iM \mybar\Phi  \Gamma \Phi\ .
  \eeq
  We are interested in matrix elements of  $ J^\mu_{A,\text{reg}} $ in a background gauge field between states without any Pauli-Villars particles, and so we need to evaluate $\vev{2iM \mybar\Phi  \Gamma \Phi}$ in a background gauge field and take the limit $M\to \infty$  to see if $\partial_\mu    J^\mu_{A,\text{reg}} $ picks up any anomalous contributions that do not decouple as we remove the cutoff.  
  
 To compute $\vev{2iM \mybar\Phi  \Gamma \Phi}$ we need to consider all Feynman diagrams with a  Pauli-Villars loop, and insertion of the $\mybar\Phi  \Gamma \Phi$ operator, and any number of external $U(1)$ gauge fields.  By gauge invariance, a graph with $n$ external photon lines will contribute $n$ powers of the field strength tensor $F^{\mu\nu}$.  For power counting, it is convenient that we normalize the gauge field so that the covariant derivative is $D_\mu=(\partial_\mu +i A_\mu)$; then the gauge field has mass dimension 1, and $F^{\mu\nu}$ has dimension 2.   In $(1+1)$ dimensions $\vev{2iM \mybar\Phi  \Gamma \Phi}$ has dimension 2, and so simple dimensional analysis implies that the graph with $n$ photon lines must make a contribution proportional to $(F^{\mu\nu})^n/M^{2(n-1)}$.   Therefore only the graph  in Fig.~\ref{fig:2Danomdiag} with one photon insertion can make a contribution that survives the $M\to\infty$ limit (the graph with zero photons vanishes).  Calculation of this diagram yields the same result for the divergence of the regulated axial current as we found in \eq{dsanomaly2}.


\begin{exercise}
\label{ex2a}
Compute the diagram in Fig.~\ref{fig:2Danomdiag} using the conventional normalization of the gauge field $D_\mu=(\partial_\mu +i e A_\mu)$ and verify that $2iM \vev{\mybar\Phi  \Gamma \Phi} = \frac{e}{2\pi}  \epsilon_{\mu\nu} F^{\mu\nu} $ when $M\to\infty$.
 \end{exercise}

Note that in this description of the anomaly we (i) effectively rendered  the  number of degrees of freedom finite by introducing the regulator; (ii) the regulator explicitly broke the chiral symmetry; (iii) as the regulator was removed, the symmetry breaking effects of the regulator never decoupled, indicating that the anomaly arises when the two vertices in Fig.~\ref{fig:2Danomdiag} sit at the same spacetime point.  While we used a Pauli-Villars regulator here, the use of a lattice regulator will have qualitatively similar features, with the inverse lattice spacing playing the role of the Pauli-Villars mass,  and we can turn these observations around:  
A lattice theory will not correctly reproduce anomalous symmetry currents in the continuum limit, unless that symmetry is broken explicitly by the lattice regulator.  This means we would be foolish to expect to construct a lattice theory with exact chiral symmetry.   But can the lattice break chiral symmetry just enough to explain the anomaly, without losing the important consequences of chiral symmetry at long distances (such as protecting fermion masses from renormalization)?

\section{ Anomalies in 3+1 dimensions}
\label{d3anomalies}

\subsection{The $U(1)_A$ anomaly}
\label{u1a}

\begin{figure}[t]
\begin{center}
  \includegraphics[width=0.30\textwidth]{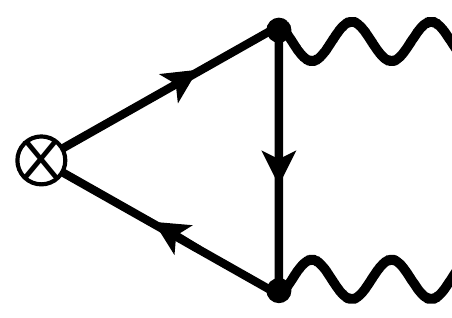}
  \end{center}
\caption{\it The $U(1)_A$ anomaly diagram in 3+1 dimensions, with one Pauli-Villars loop and an insertion of  $2iM \mybar\Phi  \Gamma \Phi$. }
\label{fig:triangle}
\end{figure}

An analogous violation of the  $U(1)_A$ current occurs in $3+1$ dimensions as well \footnote{Part of the content of this section comes directly from John Preskill's  class notes on the strong interactions, available at his web page:  {http://www.theory.caltech.edu/$\sim$preskill/notes.html}.}.  One might guess that the analogue of $\epsilon_{\mu\nu} F^{\mu\nu}=2 E$ in the anomalous divergence \eq{dsanomaly2} would be the quantity $\epsilon_{\mu\nu\rho\sigma} F^{\mu\nu}F^{\rho\sigma}=8 \vec E\cdot\vec B$, which has the right dimensions and properties under parity and time reversal.   So we should consider the behavior a massless Dirac fermion in $(3+1)$ in parallel constant $E$ and $B$ fields.   First turn on a $B$ field pointing in the $\hat z$ direction:  this gives rise to Landau levels, with energy levels $E_n$ characterized by non-negative integers $n$ as well as spin in the $\hat z$ direction $S_z$ and momentum $p_z$,  where
\beq
E_n^2 = p_z^2 + (2n+1)eB-2eB S_z\ .
\eqn{LL}\eeq
 The number density of modes per unit transverse area is defined to be $g_n$, which can be derived by computing the zero-point energy in Landau modes and requiring that it yields the free fermion result as $B\to 0$.  We have $g_n \to p_\perp d p_\perp/(2\pi)$ with $[(2n+1)eB-2eB S_z]\to p_\perp^2$, implying that 
 \beq
 g_n=eB/2\pi\ .
 \eeq
Th dispersion relation \eqn{LL} looks like that of an infinite number of  one-dimensional fermions of mass $m_{n,\pm}$, where  
\beq 
m_{n\pm}^2 =  (2n+1)eB-2eB S_z\ , \quad S_z=\pm\half \ .
\eeq
The state with $n=0$ and $S_z=+\half$ is distinguished by having $m_{n,+}=0$;  it behaves like a  {\it massless} one-dimensional Dirac fermion (with transverse density of states  $g_0$)  moving along the $\hat z$ axis  with dispersion relation $E= |p_z|$.  If we now turn on an electric field also pointing along the $\hat z$ direction  we know what to expect from our analysis in $1+1$ dimensions: we find an anomalous divergence of the axial current equal to 
\beq
g_0 e E/\pi = e^2 EB/2\pi^2
 = \left( \frac{e^2}{16\pi^2}\right)\epsilon_{\mu\nu\rho\sigma} F^{\mu\nu}F^{\rho\sigma}\ .  \eeq
 If we include an ordinary  mass term in the $3+1$ dimensional theory, then we get
 \beq
 \partial_\mu J^\mu_A = 2im\mybar\Psi\Gamma\Psi +  \left( \frac{e^2}{16\pi^2}\right)\epsilon_{\mu\nu\rho\sigma} F^{\mu\nu}F^{\rho\sigma}\ .
 \eqn{dsanomaly4}\eeq
One can derive this result by computing $\vev {M\mybar\Phi i \Gamma\Phi} $ for a Pauli-Villars regulator as in the $1+1$ dimensional example; now the relevant graph is the triangle diagram of Fig.~\ref{fig:triangle}.


If the external fields are nonabelian, the analogue of \eq{dsanomaly4} is
\beq
\partial_\mu J^\mu_A =2im\mybar\Psi\Gamma\Psi+\left( \frac{g^2}{16\pi^2}\right)\epsilon_{\mu\nu\rho\sigma}  F^{\mu\nu}_aF^{\rho\sigma}_b\Tr T_a T_b \ .  
\eqn{dsanomaly4NA}
\eeq
If the fermions transform in the defining representation of $SU(N)$, it is conventional to normalize the coupling $g$ so that $\Tr T_aT_b=\half\delta_{ab}$. This is still called an ``Abelian anomaly", since $J^\mu_A$ generates a $U(1)$ symmetry.

\subsection{Anomalies in Euclidian spacetime}
Continuing to Euclidian spacetime by means of \eqs{euc1}{euc2} changes the anomaly equations simply by eliminating the factor of $i$ from in front of the fermion mass:
\begin{align}
2d:\quad&\partial_\mu J^\mu_A =\,2m\mybar\Psi\Gamma\Psi + \frac{e}{2\pi}  \epsilon_{\mu\nu} F^{\mu\nu}\\
4d:\quad &\partial_\mu J^\mu_A =\,2m\mybar\Psi\Gamma\Psi+\left( \frac{g^2}{16\pi^2}\right)\epsilon_{\mu\nu\rho\sigma}  F^{\mu\nu}_aF^{\rho\sigma}_b\Tr T_a T_b \ .  
\eqn{eucanom4}\end{align}

\subsection{The index theorem in four dimensions}
\label{index}

  For nonabelian gauge theories  the quantity on the far right of \eq{eucanom4} is a topological charge density, with
  \beq
  \nu = \frac{g^2}{64\pi^2}\int d^4x_E\, \epsilon_{\mu\nu\rho\sigma}  F_a^{\mu\nu}F_a^{\rho\sigma}
  \eeq
   being the winding number associated with $\pi_3(G)$,  the homotopy group of maps of $S_3$ (spacetime infinity) into the gauge group $G$.
      
   Consider then continuing the anomaly equation \eq{dsanomaly4NA} to Euclidian space and integrating over spacetime its vacuum expectation value in a background gauge field (assuming the fermions to be in the $N$-dimensional representation of $SU(N)$ so that $\Tr T_aT_b=\half\delta_{ab}$).  The integral of $\partial_\mu \vev{J^\mu_A}$ vanishes because it is a pure divergence,  so we get
   \beq
  \int d^4x_E \,m\vev{\mybar\Psi \Gamma \Psi} =-\nu\ .
\eqn{ind} \eeq
The matrix element above  on the right equals
 \beq
 \int [d\Psi][d\mybar\Psi] \,e^{-S_E}\,(m\,\mybar\Psi \Gamma \Psi)\Bigl\slash\int [d\Psi][d\mybar\Psi]\, e^{-S_E} \ .
 \eeq
 where $S_E = \mybar\Psi(\Dslash_E+m)\Psi$.  We can expand $\Psi$ and $\mybar\Psi$ in terms of eigenstates of the anti-hermitian operator $\Dslash_E$, where
 \beq
 \Dslash_E \psi_n = i\lambda_n\psi_n\ ,\qquad \int d^4x_E\, \psi^\dagger_m \psi_n = \delta_{mn}\ ,
 \eeq
 with
\beq
\Psi = \sum c_n \psi_n\ ,\qquad \mybar\Psi = \sum \mybar c_n \psi^\dagger_n\ .
\eeq
Then
\beq
 \int d^4x_E\, m\,\vev{\mybar\Psi \Gamma \Psi} &=&\left(\sum_n \int d^4x_E \, m\,\psi^\dagger_n \Gamma\psi_n \prod_{k\ne n} (i\lambda_k+m)\right)\Bigl\slash\prod_k (i\lambda_k+m)\quad\cr
 &=&m \sum_n \int d^4x_E \psi^\dagger_n \Gamma\psi_n/(i\lambda_n+m) \ .
\eqn{dense5b} \eeq
    
    Recall that $\{\Gamma,\Dslash\}=0$; thus
    \beq
    \Dslash\psi_n = i\lambda_n\psi_n\quad\text{implies}\quad \Dslash(\Gamma\psi_n) = -i\lambda_n(\Gamma \psi_n)\ .
    \eeq
    Thus for   $\lambda_n\ne 0$, the eigenstates $\psi_n$ and $(\Gamma\psi_n)$ must be orthogonal to each other  (they are both eigenstates of $\Dslash$ with different eigenvalues), and so $ \psi^\dagger_n \Gamma\psi_n$ vanishes for $\lambda_n\ne 0$ and  does not contribute to the sum in \eq{dense5b}.  In contrast,
    modes with $\lambda_n=0$  can simultaneously be eigenstates of $\Dslash$ and of $\Gamma$;  let $n_+,\, n_-$ be the number of RH and LH zeromodes respectively.  The last integral in \eqn{dense5} then just equals $(n_+-n_-) = (n_R-n_L)$, and combining with \eq{ind} we arrive at the index equation
    \beq
    n_- -n_+ = \nu\ ,
    \eqn{idex}\eeq
   which states that the difference in the number of LH and RH zeromode solutions to the Euclidian Dirac equation in a background gauge field equals the winding number of the gauge field.   With $N_f$ flavors, the index equation is trivially modified to read
   \beq
     n_- -n_+ =N_f  \nu\ .
  \eqn{indexnf}   \eeq
  This link between  eigenvalues of the Dirac operator and the topological winding number of the gauge field provides a precise definition for the topological winding number of a  gauge field on the lattice (where there is no topology)  --- provided we have a definition of a lattice Dirac operator which exhibits exact zeromodes.  We will see that the overlap operator is such an operator.

\subsection{More general anomalies}
\label{naanom}

\begin{figure}[t]
\begin{center}
  \includegraphics[width=0.20\textwidth]{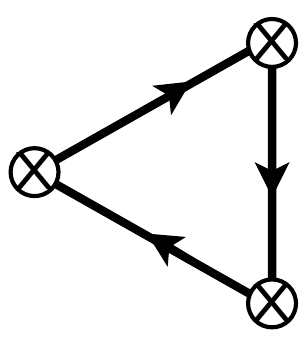}
  \end{center}
\caption{\it Anomalous three-point function of three currents.}
\label{fig:threept}
\end{figure}

   Even more generally, one can consider the 3-point correlation function of three arbitrary currents as in Fig.~\ref{fig:threept}, 
   \beq
   \vev{J^\alpha_a(k) J^\beta_b(p) J^\gamma_c(q)}
   \ ,
   \eeq
   and show that the divergence with respect to any of the indices is proportional to a particular group theory factor 
   \beq
   k_\mu  \vev{J^\mu_a(k) J^\alpha_b(p) J^\beta_c(q)}\propto  \Tr Q_a\{Q_b,Q_c\}\Bigl\vert_{R-L}\, \epsilon^{\alpha\beta\rho\sigma} k_\rho k_\sigma\ ,
   \eeq
  where the $Q$s are the generators associated with the three currents in the fermion representation,  the symmetrized trace being computed as the difference between the contributions from RH and LH fermions in the theory.   The anomaly $\CA$ for the fermion representation is defined by the group theory factor
 \beq
 \Tr \left(Q_a\{Q_b,Q_c\}\right)\Bigl\vert_{R-L} \equiv \CA\, d_{abc}\ ,
 \eqn{anomaly}  \eeq
with $d_{abc}$ being the totally symmetric invariant tensor of the symmetry group.  For a simple group $G$ (implying $G$ is not $U(1)$ and has no factor subgoups), $d_{abc}$ is only nonzero for $G=SU(N)$ with $N\ge 3$; even in the case of $SU(N)$, $d_{abc}$ will vanish for real irreducible representations (for which $Q_a=-Q_a^*$), or for judiciously chosen reducible complex  representations, such as $\bar 5\oplus 10$ in $SU(5)$.  For a semi-simple group $G_1\times G_2$ (where $G_1$ and $G_2$ are themselves simple) there are no mixed anomalies since the generators are all traceless, implying that if $Q\in G_1$ and $\CQ\in G_2$ then  $\Tr \left(Q_a\{\CQ_b,\CQ_c\}\right)\propto \Tr Q_a =0$.   When considering groups with $U(1)$ factors there can be nonzero mixed anomalies of the form $U(1)G^2$ and $U(1)^3$ where $G$ is simple;  the $U(1)^3$ anomalies can involve  different $U(1)$ groups.  With a little group theory it is not difficult to compute the contribution to the anomaly of any particular group representation.
 
 If a current with an anomalous divergence is gauged, then the theory does not make sense.  That is because the divergenceless of the current is required for the unphysical modes in the gauge field $A_\mu$ to decouple;  if they do not decouple, their propagator has a piece that goes as $k_\mu k_\nu/k^2$ which does not fall off at large momentum, and the theory is not renormalizable.
 
 When global $U(1)$ currents have anomalous divergences, that is interesting.  We have seen that the $U(1)_A$ current is anomalous, which explains the $\eta'$ mass;  the divergence of the axial isospin current explains the decay $\pi^0\to\gamma\gamma$;  the anomalous divergence of the baron number current in background  $SU(2)$ in the Standard Model predicts baryon violation in the early universe and the possibility of weak-scale baryogenesis.
 %

\begin{exercisebn}
\label{ex2b}
Verify that all the gauge currents are anomaly-free in the standard model with the representation in \eq{SM}.  The only possible $G^3$ anomalies are for $G=SU(3)$ or $G=U(1)$; for the $SU(3)^3$ anomaly use the fact that a LH Weyl fermion contributes $+1$ to $\CA$ if it transforms as a $3$ of $SU(3)$, and contributes $-1$ to $\CA$ if it is a $\mybar 3$. There are two mixed anomalies to check as well: $U(1)SU(2)^2$ and $U(1)SU(3)^2$.

This apparently miraculous cancellation is suggestive that each family of fermions may be unified into a spinor of $SO(10)$, since the vanishing of anomalies which happens automatically in $SO(10)$ is of course maintained when the symmetry is broken to a smaller subgroup, such as the Standard Model.

\end{exercisebn}
 \begin{exercisenb}
\label{ex2c}
Show that the global $B$ (baryon number) and $L$ (lepton number) currents are anomalous in the Standard Model \eq{SM}, but that $B-L$ is not.
 \end{exercisenb}

\section{Strongly coupled chiral gauge theories}

Strongly coupled chiral gauge theories are particularly intriguing, since they can contain light composite fermions, which could possibly describe  the quarks and leptons we see. A nice toy example of a strongly coupled chiral gauge theory is $SU(5)$ with LH fermions  
\beq
\psi =\mybar 5\ ,\qquad \chi= 10\ .
\eeq 
It so happens that the $\psi$ and the $\chi$ contribute with opposite signs to the $\text{(SU(5))}^3$ anomaly $\CA$ in \eq{anomaly}, so this seems to be a well defined gauge theory.  Furthermore, the $SU(5)$  gauge interactions are asymptotically free, meaning that interactions becomes strong at long distances.  One might therefore expect the theory to confine as QCD does.  However, unlike QCD, there are no gauge invariant fermion bilinear condensates which could form, and which in QCD are responsible for baryon masses.   That being the case, might there be any massless composite fermions in the spectrum of this theory? 't Hooft came up with a nice general argument  involving global anomalies  that suggests there will be.

In principle there are two global $U(1)$ chiral symmetries in this theory corresponding to independent phase rotations for $\psi$ and $\chi$; however both of these rotations have $\text{global}\times \text{SU(5)}^2$ anomalies, similar to the $\text{global}\times \text{SU(3)}^2$ of the $U(1)_A$ current in QCD.  This anomaly can only break one linear combination of the two $U(1)$ symmetries, and one can choose the orthogonal linear combination  which is anomaly-free.  With a little group theory you can show that the anomaly-free global $U(1)$ symmetry corresponds to assigning charges
\beq
\psi =\mybar 5_3\ ,\qquad \chi= 10_{-1}\ ,
\eeq
where the subscript gives the global $U(1)$ charge. This theory has a nontrivial global $U(1)^3$ anomaly, $\CA = 5\times (3)^3 + 10 \times (-1)^3 = 125$.   't Hooft's argument is that this$\text{(global)}^3$  anomaly restricts --- and helps predict --- the low energy spectrum of the theory.  Applied to the present model, his argument goes as follows:  imagine weakly gauging this $U(1)$ symmetry.  This would be bad news as the theory stands, since a $\text{(gauge)}^3$ anomaly leads to a sick theory, but one can add a LH ``spectator fermion'' $\omega=1_{-5}$ which is a singlet under $SU(5)$  but has charge $-5$ under this $U(1)$ symmetry, canceling the $U(1)^3$ anomaly.  This weak $U(1)$ gauge interaction plus the SU(5)-singlet $\omega$ fermion should not interfere with the strong $SU(5)$ dynamics.  If that dynamics leads to confinement and no $U(1)$ symmetry breaking, then the weak $U(1)$ gauge theory must remain anomaly free at low energy, implying that there has to be one or more massless composite fermions to cancel the $U(1)^3$ anomaly of the $\omega$.  A good candidate  massless composite LH fermion is $(\psi\psi\chi)$ which is an $SU(5)$-singlet (as required by confinement), and which has $U(1)$ charge of $(3+3-1)=5$, exactly canceling the $U(1)^3$ anomaly of the $\omega$.  Now forget the thought experiment: do not gauge the $U(1)$ and do not include the $\omega$ spectator fermion.  It should still be true that this $SU(5)$ gauge theory produces a single massless composite fermion $(\psi\psi\chi)$ \footnote{You may wonder about whether fermion condensates form which break the global $U(1)$ symmetry.  Perhaps, but it seems unlikely.  The lowest dimension gauge invariant fermion condensates involve four fermion fields --- such as  $\vev{\chi\chi\chi\psi}$ or  $\vev{(\chi\psi)(\chi\psi)^\dagger}$ --- which are all neutral   under the $U(1)$ symmetry.  Furthermore, there are arguments that a Higgs phase would not be distinguishable from a confining phase for this theory.}.

While it is hard to pin down the spectrum of  general strongly coupled chiral gauge theories using 't Hooft's anomaly matching condition alone, a lot is known about 
 strongly coupled supersymmetric chiral gauge theories, and they typically have a very interesting spectrum of massless composite fermions, which can be given small masses and approximate the quarks and leptons we see by tweaking the theory.  See for example \cite{Kaplan:1997tu}, which constructs a theory with three families of massless composite fermions, each with a different number of constituents.  
 
 Chiral gauge theories would be very interesting to study on the lattice, but pose theoretical problems that have not been solved yet--- and which if they were, might then be followed by challenging practical problems related to complex path integral measures and massless fermions.   Perhaps one of you will crack this interesting problem.

\section{The non-decoupling of parity violation in odd dimensions}
\label{parityanomalies}

We have seen that chiral symmetry does not exist in odd space dimensions, but that a discrete parity symmetry can forbid a fermion mass.  One would then expect a regulator --- such as Pauli-Villars fields --- to break parity. Indeed they do: on integrating the Pauli-Villars field out of the theory, one is left with a Chern Simons term in the Lagrangian with coefficient $M/|M|$, which does not decouple as $M\to\infty$. In $2k+1$ dimensions the Chern Simons form for an Abelian gauge field is proportional to
\beq
\epsilon^{\alpha_1\cdots\alpha_{2k+1} } A_{\alpha_1}F_{\alpha_2\alpha_3}\cdots F_{\alpha_{2k}\alpha_{2k+1}}
\eeq
which violates parity; the Chern Simons form for nonabelian gauge fields is more complicated.

\begin{figure}[t]
\begin{center}
  \includegraphics[width=0.40\textwidth]{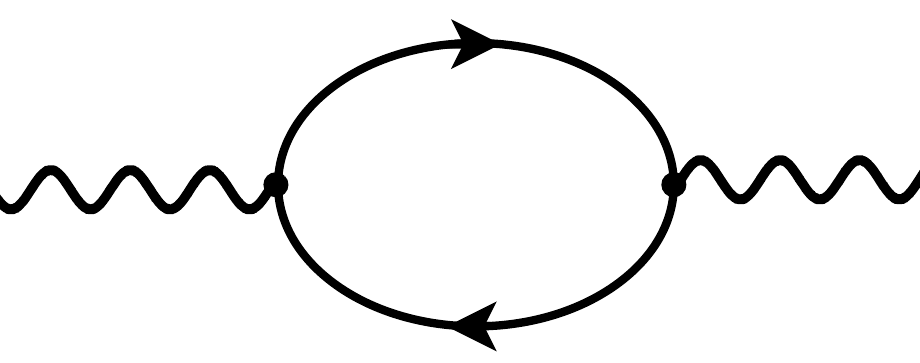}
  \end{center}
\caption{\it Integrating out a heavy fermion in three dimensions gives rise to the Chern Simons term in the effective action of \eq{csact}.}
\label{fig:CS}
\end{figure}

For domain wall fermions we will be interested in a closely related but slightly different problem:  the generation of a Chern Simons operator on integrating out a heavy fermion of mass $m$.  In $1+1$ dimensions with an Abelian gauge field one computes the graph in Fig.~\ref{fig:CS}, which gives rise to the Lagrangian
\beq
\CL_{CS} = \frac{e^2}{8\pi}\frac{m}{|m|} \epsilon^{\alpha\beta\gamma} A_\alpha \partial_\beta A_\gamma\ .
\eqn{csact}\eeq
What is interesting is that it implies a particle number current
\beq
J_\mu = \frac{1}{e}\frac{\partial \CL_{CS}}{\partial A_\mu} = \frac{e}{8\pi} \frac{m}{|m|} \epsilon^{\mu\alpha\beta} F_{\alpha\beta}
\eeq
which we will see is related to the anomaly \eq{dsanomaly2} in $1+1$ dimensions.

 \begin{exercise}
\label{ex2d}
Verify the coefficient in \eq{csact} by computing the diagram Fig.~\ref{fig:CS}.  By isolating the part that is proportional to  $\epsilon_{\mu\nu\alpha} p^\alpha$ before performing the integral, one can make the diagram very easy to compute. 
\end{exercise}

%
%
\chapter{Domain Wall Fermions}

\section{Chirality, anomalies and fermion doubling}

You have heard of the Nielsen-Ninomiya theorem: it states that a  fermion action in $2k$ Euclidian spacetime dimensions
\beq
S=\int_{\pi/a}^{\pi/a}  \frac{d^{2k}p}{(2\pi)^4}  \, \mybar\Psi_{-\bfp} \tilde D(\bfp) \Psi(\bfp)
\eeq
cannot have the operator $\tilde D$ satisfy all four of the following conditions simultaneously:
\begin{enumerate}
\item $\tilde D(\bfp)$ is a periodic, analytic function of $p_\mu$;
\item $D(\bfp)\propto \gamma_\mu p_\mu$ for $a |p_\mu|\ll 1$;
\item $\tilde D(\bfp)$ invertible everywhere except $p_\mu=0$;
\item $\{\Gamma, \tilde D(\bfp)\}=0$.
\end{enumerate}

The first condition is required for locality of the Fourier transform of $\tilde D(\bfp)$ in coordinate space.  The next two state that we want a single flavor of conventional Dirac fermion in the continuum limit.  The last item is the statement of chiral symmetry.  One can try keeping that and eliminating one or more of the other conditions;  for example, the SLAC derivative took $\tilde D(\bfp) = \gamma_\mu p_\mu$ within the Brillouin zone (BZ), which violates the first condition --- if taken to be periodic, it is discontinuous at the edge of the BZ.  This causes problems ---  for example, the QED Ward identity states that the photon vertex $\Gamma_\mu$ is proportional to $\partial\tilde D(\bfp)/\partial p_\mu$, which is infinite at the BZ boundary.  Naive fermions satisfy all the conditions except  (3):  there $\tilde D(\bfp)$ vanishes at the $2^4$ corners of the BZ, and so we have $2^4$ flavors of Dirac fermions in the continuum. Staggered fermions are somewhat less redundant, producing four flavors in the continuum for each lattice field;  Creutz fermions are the least redundant, giving rise to two copies for each lattice field.  The discussion in any even spacetime dimension is analogous.

This roadblock in developing a lattice theory with chirality is obviously impossible to get around when you consider anomalies.  Remember that anomalies do occur in the continuum but that in a UV cutoff on  the number of degrees of freedom, there are no anomalies, and the exact symmetries of the regulated action are the exact symmetries of the quantum theory.  The only way a  symmetry current can have a nonzero divergence is if either the original action or the UV regulator explicitly violate that symmetry.  The implication for lattice fermions is that any symmetry that is exact on the lattice will be exact in the continuum limit, while any symmetry anomalous in the continuum limit must be broken explicitly on the lattice.

A simple example to analyze is the case of a ``naive''  lattice action for a single RH fermion,
\beq
S &=&\frac{1}{2a} \sum_{\bfn, \mu} \,\mybar\Psi_+(\bfn) \gamma_\mu \left[\Psi_+(\bfn +\hat{\mathbf \mu}) - \Psi_+(\bfn -\hat{\mathbf \mu}) \right]\cr
&=&\frac{1}{2a}\sum_{\bfp, \mu} 2i\sin ap_\mu \mybar\Psi_+(-\bfp)\gamma_\mu \Psi_+(\bfp)\ ,\cr \Gamma \Psi_+& =& \Psi_+,
\eeq
so that $\tilde D(\bfp) = i\gamma_\mu \sin a k_\mu/a$. This vanishes at every corner of the BZ; expanding about these points we write  $p_\mu =q_\mu+ n_\mu \pi/a$ with $n_\mu\in \{0,1\}$ and $a q_\mu\ll 1$ and find 
\beq
\tilde D(\bfp) \simeq i \sum_\mu(-1)^{n_\mu}\gamma_\mu q_\mu\ .
\eqn{bzd}
\eeq
  These zeroes of $\tilde D(\bfp)$ (at $q_\mu=0$) correspond to the $2^d$ doublers in $d$-dimensional Euclidian spacetime, violating condition (3) in the Nielsen-Ninomiya theorem.  However,  $\tilde D(\bfp)$ does satisfy condition (4) and the action $S$  is  invariant under the  symmetry
  \beq
  \Psi_+(\bfp)\to e^{i\alpha}\Psi_+(\bfp)=e^{i\alpha \Gamma}\Psi_+(\bfp)\ ,
  \eeq
which looks like a chiral symmetry --- yet for a continuum theory with $2^d$ RH fermions, a phase symmetry would be anomalous, which we know cannot result from a symmetric lattice theory! 

The resolution is that the continuum theory does {\it not }have $2^d$ RH fermions, but rather $2^{d-1}$ Dirac fermions, and the exact lattice $U(1)$ symmetry corresponds to an exact fermion number symmetry in the continuum, which is not chiral and not anomalous.
To show this, note that $\tilde D(\bfp)$ in \eq{bzd} has funny signs near the corners of the BZ.  We can convert back to our standard gamma matrix basis using the similarity transformation  $P(n)\gamma_\mu P(n)^{-1} = (-1)^{n_\mu}\gamma_\mu$;  but then $P(n)\Gamma P(n)^{-1} = (-1)^{\sigma(n)}\Gamma$, where $\sigma(n) = \sum_\mu n_\mu$ (since $\Gamma$ is the product of all the $\gamma^\mu$).  Therefore
 \beq
 \Gamma [P(n)\Psi_+(\bfq)]=(-1)^{\sigma(n)} [P(n)\Psi_+(\bfq)]
 \eqn{altchi}\eeq
  and in the continuum we have $2^{d-1}$ RH fermions and $2^{d-1}$ LH fermions, and the exact and apparently chiral symmetry of the lattice corresponds to an exact and anomaly-free fermion number symmetry in the continuum. The redundancy of staggered and Creutz fermions serve the same purpose, ensuring that all exact lattice symmetries become anomaly-free vector symmetries in the continuum limit.

\section{Domain wall fermions in the continuum}
\label{dwf}
\subsection{Motivation}

What we would like is a realization of chiral symmetry on the lattice which (i) is not exact, so we can correctly recover anomalies, but which (ii) retains all the good features of chiral symmetry in the continuum, such as protection from additive renormalization of fermion masses.  There is a curious example in the continuum of such a system, which gave a clue on how to achieve this.  The example has to do with fermions in odd dimension that interact with a domain wall.  To be concrete, consider a system in three dimensions (coordinates $(x_0,x_1,x_2)$), where the fermion has a mass which depends on $x_2$ and switches sign at $x_2=0$.  For simplicity, I will take $m(x_2)=m\epsilon(x_2)= m x_2/|x_2|$.
Curiously enough, we will show that a massless fermion mode exists bound to this 2-dimensional surface, and that it is chiral:  there exist a RH mode, and not a LH one.  Thus the low energy limit of this theory looks like a 2-dimensional theory of a Weyl fermion with a chiral symmetry, even though we started with a 3-dimensional theory in which  there can be no chirality.

Yet we know that the low energy effective theory is anomalous.  Recall that for a massless Dirac fermion coupled to photons in two Euclidian dimensions, the vector current is conserved ( $\partial_\mu J^\mu = 0$),  while the axial current is not in general  ($\partial_\mu J^\mu_A = (e^2/\pi E)$).  Thus the fermion current for a RH Weyl current satisfies
\beq
\partial_\mu J^\mu_R = \half\partial_\mu (J^\mu  J^\mu_A) = \frac{e}{2\pi} E\ .
\eeq
If we turn on an electric field pointing in the $x_1$ direction, then the charge on the mass defect must increase with time.  Yet in the full 3-dimensional theory, there is only one fermion current $J^\mu = \mybar\Psi \gamma^\mu\Psi$, and it is conserved.  So even though only massive states live off the mass defect, we see that they must somehow know about the anomaly and allow chiral symmetry to be violated as the anomaly requires.

You might be suspicious that there is some hidden fine tuning here to keep the chiral mode massless, but that cannot be:  the low energy effective theory would need a LH mode as well in order for there to be a mass.  Distortions of the domain wall mass function cannot change this result, unless the mass $m(x_2)$ becomes small enough somewhere to change the spectrum of the low energy effective theory.  But even in that case there is an index theorem that requires there to be a massless Weyl fermion as the sign of $m(x_2)$ changes at an odd number  of locations.

This looks like a useful trick to apply to the lattice:  an anomalous chiral symmetry emerging at low energy from a full theory with no fundamental chiral symmetry, and without fine tuning.  In this lecture I show how the continuum theory works, and then how it can be transcribed to the lattice.  In the next lecture I will discuss how the effective theory can be described directly using the overlap formulation, without any reference to the higher dimensional parent theory.

\subsection{The model}

Even though fermions in even and odd dimensions look quite different, one finds an interesting connection between them when considering the Dirac equation with a space-dependent mass term.  One can think of a space-dependent mass as arising from a Higgs mechanism, for example, where there is a topological defect trapped in the classical Higgs field, such as a domain wall or a vortex.  A domain wall can naturally arise when the Higgs field breaks a discrete symmetry;  a vortex when the Higgs field breaks a $U(1)$ symmetry (see John Preskill's lectures ``Vortices and Monopoles" at the 1985 Les Houches Summer School \cite{Preskill:1986kp}.  Domain wall defects are pertinent to putting chiral fermions on the lattice, so I will consider that example.

Consider a  fermion in Euclidian spacetime with  dimension $d=2k+1$, where the coordinates are written as $\{x_0,x_1,\ldots x_{2k-1},s\}\equiv\{x_\mu,s\}$, where $\mu=0,\ldots,2k-1$ and $s$ is what I call the  coordinate $x_{2k}$.   The $(2k+1)$ $\gamma$ matrices are written as $\{\gamma_0,\ldots,\gamma_{2k},\Gamma\}$. This fermion is assumed to have an $s$-dependent mass  with the simple form
\beq
m(s) = m \epsilon(s) = \begin{cases} +m & s>0\cr -m & s<0\end{cases}\ ,\qquad m>0\ .
\eqn{massfunc}
\eeq
This mass function
explicitly breaks the Poincar\'e symmetry of $2k+1$ dimensional spacetime, but preserves the Euclidian Poincar\'e symmetry of $2k$ dimensional spacetime.  The fermion is also assumed to interact with $2k$-dimensional  background gauge fields $A_\mu(x_\mu)$ which are independent of $s$. The   Dirac equation may be written as:
\beq
\left[\Dslash + \Gamma\partial_s+m(s)\right]\Psi(x_\mu,s)=0\ ,
\eqn{dwdirac}
\eeq
where $\Dslash$ is the lower dimension ($d=2k$) covariant Dirac operator.
The spinor $\Psi$ can be factorized as the product of a functions of $s$ times spinors $\psi(x_\mu)$, 
\beq
\Psi(x_\mu,s)= \sum_{n} \left[{
b_n(s)}{P_+} +{ f_n(s)}
{P_-}\right] \psi_n(x_\mu)\ ,\qquad P_\pm = \frac{1\pm \Gamma}{2}\ ,
\eeq
satisfying the equations
\beq
[\partial_s + m(s)]{b_n(s)} &=&{\mu_n} {f_n(s)}\ ,\cr [-\partial_s + m(s)] {f_n(s)} 
&= &
{\mu_n} {b_n(s)}\ ,
\eqn{susy}\eeq
and
\beq
( \Dslash+\mu_n)\Psi_n(x) &=& 0\ .
\eqn{deq}
\eeq
One might expect all the eigenvalues in \eq{susy} to satisfy $|\mu_n|\gtrsim O(m)$, since that is the only scale in the problem.  However, there is also a  solution to \eq{susy} with eigenvalue  $\mu=0$ given by
\beq
b_0 = Ne^{-\int^s_0 m(s') ds'} = N e^{-m|s|}\ .
\eqn{zmode}\eeq
This solution is localized near the defect at $s=0$, falling off exponentially fast away from it.  There is no analogous solution to \eq{susy} of the form $$f_0\sim e^{+\int^s_0 m(s') ds'}\ ,$$since that would be exponentially growing in $|s|$ and not normalizable.    Therefore as seen from \eq{deq} the spectrum consists of an infinite tower of fermions satisfying the  $d=2k$ Dirac equation:  massive Dirac fermions with mass $O(m)$ and higher, plus a single massless right-handed chiral fermion.  The massless fermion is localized at the defect at $s=0$, whose profile in the transverse extra dimension is given by \eq{zmode}; the massive fermions are not localized.  Because of the gap in the spectrum, at low energy the accessible part of the spectrum consists only of the massless RH chiral fermion.

\begin{exercise}
\label{ex1c}
Construct a $d=2k+1$ theory whose low energy spectrum possesses a single light $d=2k$ Dirac fermion with mass arbitrarily lighter than the domain wall scale $m$. There is more than one way to do this.
 \end{exercise}

Some comments are in order:
\begin{itemize}
\item
It is not a problem that the low energy theory of a single right-handed chiral fermion violates parity in $d=2k$ since the mass for $\Psi$ breaks parity in $d=2k+1$;
\item
Furthermore, nothing is special about right-handed fermions, and a left handed mode would have resulted if we had chosen the opposite sign for the mass in \eq{massfunc}.   This makes sense because choosing the opposite sign for the mass can be attained by flipping the sign of all the space coordinates: a rotation in the $(2k+1)$ dimensional theory, but a parity transformation from the point of view of the $2k$-dimensional fermion zeromode.
\item
The fact that a chiral mode appeared at all is a consequence of the normalizability of  $\text{exp}(-\int^s_0 m(s') ds')$, which in turn follows from the two limits $m(\pm\infty)$  being nonzero with opposite signs.   Any function $m(s)$ with that boundary condition will support a single chiral mode, although in general there may also be a number of very light fermions localized in regions wherever $|m(s)|$ is small --- possibly extremely light if $m(s)$ crosses zero a number of times, so that there are widely separated defects and anti-defects.  
\item
Gauge boson loops will generate contributions to the fermion mass function which are even in $s$.  If the coupling is sufficiently weak, it cannot effect the masslessness of the chiral mode.  However if the gauge coupling is strong, or if the mass $m$ is much below the cutoff of the theory, the radiative corrections could cause the fermion mass function to never change sign, and the chiral mode would not exist.  Or it could still change sign, but become small in magnitude in places, causing the chiral mode to significantly delocalize.  An effect like this can cause trouble with lattice simulations at finite volume and lattice spacing; more later.
\end{itemize}

%
 
\section{Domain wall fermions and the Callan-Harvey mechanism}
\label{CallanHarvey}

Now turn on the gauge fields and see how the anomaly works, following  \cite{Callan:1984sa}.  To do this, I integrate out the heavy modes in the presence of a background gauge field.  Although I will be interested in having purely $2k$-dimensional gauge fields in the theory, I will for now let them be arbitrary $2k+1$ dimensional fields.  And since it is hard to integrate out the heavy modes exactly, I will assume perform the calculation as if their mass was constant, and then substitute $m(s)$;  this is not valid where $m(s)$ is changing rapidly (near the domain wall) but should be adequate farther away.  Also --- in departure from the work of \cite{Callan:1984sa}, I will include a Pauli-Villars field with constant mass $M<0$, independent of $s$;  this is necessary to regulate fermion loops in the wave function renormalization for the gauge fields, for example.  

When one integrates out the heavy fields, one generates a Chern Simons operator in the effective Lagrangian, as discussed in \S\ref{parityanomalies}:
\beq
\CL_\text{CS} = \left(\frac{m(s)}{|m(s)|}+ \frac{M}{|M|}\right) \CO_{CS} = \left(\epsilon(s) -1\right) \CO_{CS}  
\eeq
Note that with $M<0$,  the coefficient of the operator equals $-2$ on the side where $m(s)$ is negative, and equals zero on the side where it is positive. For a background $U(1)$ gauge field one finds in Euclidian spacetime:
\begin{align}
d=3:\quad &\CO_{CS} = -\frac{e^2}{8\pi}\epsilon_{abc}( A_a \partial_a A_c)\ ,\\
d=5:\quad &\CO_{CS} = -\frac{e^3}{48\pi^2}\epsilon_{abcde}( A_a \partial_b A_c \partial_d A_e)\ .
\end{align}
Differentiating $\CL_\text{CS} $ by $A_\mu$ and dividing by $e$ gives the particle number current:
\beq
J^{(CS)}_a = \left(\epsilon(s) -1\right) \begin{cases}  -\frac{e}{8\pi}\epsilon_{abc}(F_{bc})&d=3\cr &\cr
 -\frac{e^2}{64\pi^2}\epsilon_{abcde}( F_{bc}F_{de})&d=5
\end{cases}
\eqn{JCS} \eeq
where I use Latin letters to denote the coordinates in $2k+1$ dimensions, while Greek letters will refer to indices on the $2k$-dimensional defect.
So when we turn on background $2k$ dimensional gauge fields, particle current flows either onto or off of the domain wall along the transverse $s$ direction on the left side (where $m(s) = -m$).  If we had regulated with a positive mass Pauli Villars field, the current would flow on the right side.  But in either case, this bizarre current exactly accounts for the anomaly.  Consider the case of a 2-dimensional domain wall embedded in 3-dimensions.  If we turn on an $E$ field we know that from the point of view of a 2d creature, RH Weyl particles are created, where from \eq{eucanom4},  
\beq
\partial_\mu J_{\mu,R} = \half\partial_\mu J_{\mu,A} = \frac{e}{4\pi} \epsilon_{\mu\nu} F_{\mu\nu}\ .
\eeq
We see from \eq{JCS} this current is exactly compensated for by the Chern Simons current $J^{(CS)}_2 = \frac{e}{4\pi} \epsilon_{2\mu\nu} F_{\mu\nu}$ which flows onto the domain wall from the $-s = -x_2$ side.  The total particle current is divergenceless.  

This is encouraging:  (i) we managed to obtain a fermion whose mass is zero due to topology and not fine tuning;  (ii) the low energy theory therefore has a chiral symmetry even though the full 3d theory does not;  (iii) the only remnant of the explicit chiral symmetry breaking of the full theory is the anomalous divergence of the chiral symmetry in the presence of gauge fields.  One drawback though is the infinite dimension in the $s$ direction, since we will eventually want to simulate this on a finite lattice; besides, it is always disturbing to see currents streaming in from $s=-\infty$!  One solution is to work in finite $(2k+1)$ dimensions, in which case we end up with a massless RH mode stuck to the boundary on one side and a LH mode on the other (which is great for a vector like theory of massless Dirac fermions, but not for chiral gauge theories).  This is what one does when simulating domain wall fermions.  The other solution is more devious, leads to the ``overlap operator", and is the subject of another day's lecture.

\subsection{Domain wall fermions on a slab}

To get a better understanding for how the theory works, it is useful to consider a compact extra dimension. In particular, consider the case of periodic boundary conditions $\Psi(x_\mu,s+2s_0) = \Psi(x_\mu,s)$;  we define the theory on the interval $-s_0\le s \le s_0$ with $\Psi(x_\mu,-s_0) = \Psi(x_\mu, s_0)$ and mass $m(s) = m\frac{s}{|s|}$.  Note the the mass function $m(s)$ now has a domain wall kink at $s=0$ and an anti-kink at $s=\pm s_0$.  There are now two exact zeromode solutions to the Dirac equation,
\beq
b_0(s) = N e^{-\int_{-s_0}^s m(s') ds'}\ ,\qquad
f_0(s)= N e^{+\int_{-s_0}^s m(s') ds'}\ .
\eqn{zmodes}
\eeq
Both solutions are  normalizable since the transverse direction is finite;  $b_0$ corresponds to a right-handed chiral fermion located at $s=0$, and $f_0$ corresponds to a left-handed chiral fermion located at $s=\pm s_0$.  However, in this case the existence of exactly massless modes is a result of the fact that $\int_{-s_0}^{+s_0} m(s)\, ds=0$ which is not a topological condition and not robust.  For example, turning on weakly coupled gauge interactions will cause a shift the mass by $\delta m(s)\propto \alpha m$ (assuming $m$ is the cutoff) which ruins this property.   However: remember that to get a mass in the $2k$-dimensional defect theory, the RH and LH chiral modes have to couple to each other.  The induced mass will be  
\beq
\delta\mu_0 \sim \delta m\, \int ds\, b_0(s) \,f_0(s) = \delta m\,N^2 \sim\alpha m  \times \frac{2m s_0}{\cosh[m s_0]}\equiv m_\text{res}
\eqn{mres}\eeq
which vanishes exponential fast as $(M s_0)\to\infty$. Nevertheless, at finite $s_0$ there will always be some chiral symmetry breaking, in the form of a residual mass, called $m_{res}$.  
If, however, one wants to work on a finite line segment in the extra dimension instead of a circle, we can take an asymmetric mass function,
\beq
m(s) = \begin{cases} -m & -s_0\le s\le 0\cr +\infty &  0<s<s_0 \end{cases}
\eeq
This has the effect of excluding half the space, so that the extra dimension has boundaries at $s=-s_0$ and $s=0$.  Now even without extra interactions, one finds
\beq
m_\text{res} \sim 2m e^{-2ms_0}
\eeq

Any matrix element of a chiral symmetry violating operator will be proportional to the overlap of the LH and RH zeromode wave functions, which is proportional to $m_\text{res}$.
On the lattice the story of $m_\text{res}$ is more complicated ---  as  discussed in \S3.4 --- both because of the discretization of the fermion action, and because of the presence of rough gauge fields.   Lattice computations with domain wall fermions need to balance the cost of simulating a large extra dimension versus the need to make $m_{res}$ small enough to attain chiral symmetry.

\subsection{The (almost) chiral propagator}
Before moving to the lattice, I want to mention an illuminating calculation by L\"uscher \cite{Luscher:2000hn} who considered  noninteracting domain wall fermions with a semi-infinite fifth dimension, negative fermion mass, and  and LH Weyl fermion zeromode bound to the boundary at $s=0$.  He computed the Green function for propagation of the zeromode from $(x,s=0)$ to $(y,s=0)$   and examined the chiral properties of this propagator.   The differential operator to invert  should be familiar now:
\beq
D_5 =\dslash_4 + \gamma_5\partial_s -m\ , \qquad s\ge 0\ .
\eeq
We wish to look at the Green function
 $G$ which satisfies
\beq
D_5 G(x,s;y,t) = \delta^{(4)}(x-y) \delta(s-t)\ ,\qquad P_+ G(x,0;y,t)=0\ .
\eeq
The solution L\"uscher found for propagation along the boundary was
\beq
G(x,s; y,t) \biggl\vert_{s=t=0} =2  P_- D^{-1} P_+\ ,
\eqn{gsol}\eeq
where $D$ is the peculiar looking operator
\beq
D=\left[ 1+\gamma_5 \epsilon(H)\right]\ ,\qquad H \equiv \gamma_5(\dslash_4-m)=H^\dagger\ ,\qquad \epsilon(\CO)\equiv \frac{\CO}{\sqrt{\CO^\dagger\CO}}\ .
\eqn{lresult}\eeq
This looks pretty bizarre!  Since $H$ is hermitian, in a basis where $H$ is diagonal, $\epsilon(H)=\pm 1$!  But don't conclude that in this basis the operator is simply $D=(1\pm \gamma_5)$ --- you must remember, that in the basis where $H$ is diagonal, $\epsilon(H)\gamma_5$ is not  (by which I mean $\expect{m}{\epsilon(H)\gamma_5}{n}$ is in general nonzero for $m\ne n$ in the $H$ eigenstate basis).
In fact, \eq{lresult} looks very much like the overlap operator discovered some years earlier and which we will be discussing soon.  

A normal Weyl fermion in four dimensions would have a propagator $P_- (\dslash_4)^{-1} P_+$; here we see that the domain wall fermion propagator looks like the analogous object arising from the fermion  action $\mybar\Psi D\Psi$, with $D$  playing the role of the four-dimensional Dirac operator $\dslash_4$.  So what are the properties of $D$?
\begin{itemize}
\item For long wavelength modes (e.g. $k\ll m$) we can expand $D$ in powers of $\dslash_4$ and find
\beq
D=\frac{1}{m}\left(\dslash_4 - \frac{\partial_4^2}{2m} + \ldots\right)\ ,
\eeq
which is reassuring:  we knew that at long wavelengths we had a garden variety Weyl fermion living on the boundary of the extra dimension (the factor of $1/m$ is an unimportant normalization).  
\item
A massless Dirac action is chirally invariant because $\{\gamma_5,\dslash_4\}=0$.  However, the operator $D$ does not satisfy this relationship, but rather:
\beq
\{\gamma_5,D\} =  D\gamma_5 D\ ,
\eqn{gw1}\eeq
or equivalently,
\beq
\{\gamma_5,D^{-1}\} =  \gamma_5 \ .
\eqn{gw2}
\eeq
This is the famous Ginsparg-Wilson equation, first introduced  in context of the lattice (but not solved) many years ago \cite{Ginsparg:1981bj}.  Note  the right hand side of the above equations encodes the violation of chiral symmetry that our Weyl fermion experience; the fact that the right side of \eq{gw2} is local in spacetime implies that violations of chiral symmetry will be seen in Green functions {\it only} when operators are sitting at the same spacetime point.  We know from our previous discussion, the only chiral symmetry violation that survives to low energy in the domain wall model is the anomaly, and so it must be that the chiral symmetry violation in \eqs{gw1}{gw2} encode the anomaly and nothing else, at low energy \footnote{A lattice solution to \eq{gw1} (the only solution in existence)  is the overlap operator discovered  by Neuberger \cite{Neuberger:1997fp,Neuberger:1998wv}; it was a key reformulation of earlier work  \cite{Narayanan:1993ss,Narayanan:1994gw} on how to represent domain wall fermions with an infinite extra dimension (and therefore exact chiral symmetry) in terms of entirely lower dimensional variables.  We will discuss overlap fermions and the Ginsparg-Wilson equation further in the next lecture.}.
\end{itemize}

\section{Domain wall fermions on the lattice}
\label{DWFL}

The next step is to transcribe this theory onto the lattice.  If you replace continuum derivatives with the usual lattice operator $D\to \half(\nabla^* + \nabla)$ (where $\nabla$ and  $\nabla^*$ are the forward and backward lattice difference operators respectively) then one discovers...doublers!  Not only are the chiral modes doubled in the $2k$ dimensions along the domain wall, but there are two solutions for the transverse wave function of the zero mode, $b_0(s)$, one of which alternates sign with every step in the $s$ direction and which is a LH mode.  So this ends up giving us a theory of naive fermions on the lattice, only in a much more complicated and expensive way!

However, when we add Wilson terms $\frac{r}{2} \nabla^*\nabla$ for each of the dimensions, things get interesting.  You can think of these as mass terms which are independent of $s$ but which are dependent on the wave number $k$ of the mode, vanishing for long wavelength.  What happens of we add a $k$-dependent spatially constant mass $\Delta m(k)$  to the step function mass $m(s)=m\epsilon(s)$?  The solution for $b_0(s)$ in \eq{zmode} for an infinite extra dimension becomes
\beq
b_0= N e^{-\int^s_0 [m(s') +\Delta m(k)]ds'}\ ,
\eeq
which is a normalizable zeromode solution --- albeit, distorted in shape --- so long as $|\Delta m(k)| < m$.  However, for $|\Delta m(k)| > m$, the chiral mode vanishes.  What happens to it?  It becomes more and more extended in the extra dimension until it ceases to be normalizable. What is going on is easier to grasp for a finite extra dimension:  as $ |\Delta m(k)|$ increases with increasing $k$, eventually the $b_0$ zeromode solution extends to the opposing boundary of the extra dimension, when $|\Delta m(k)| \sim (m-1/s_0)$.  At that point it can pair up with the $LH$ mode and become heavy.


\begin{figure}[t]
\begin{center}
  \includegraphics[width=0.85\textwidth]{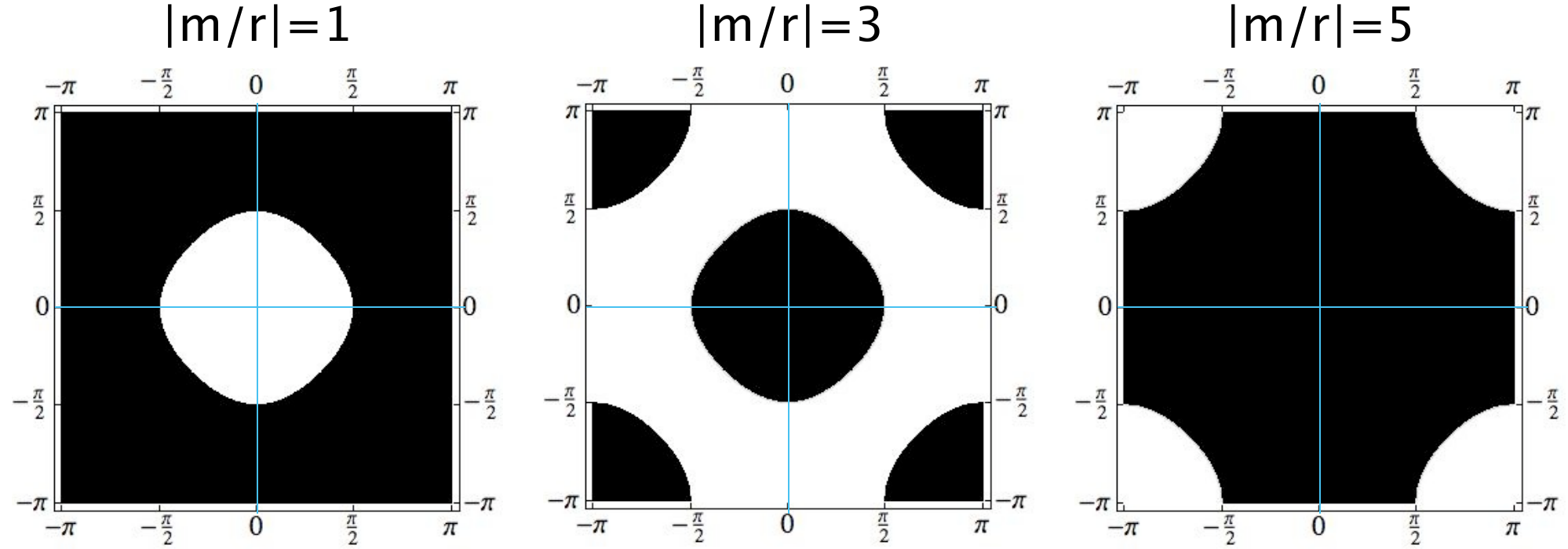}
  \end{center}
\caption{\it Domain wall fermions in $d=2$ on the lattice:  dispersion relation plotted in the Brillouin zone.  Chiral modes exist in white regions only.  For $0<|m/r|<2$ there exists a single RH mode centered at $(k_1,k_2)=(0,0)$. for $2<|m/r|<4$ there exist two LH modes centered at $(k_1,k_2)=(\pi,0)$ and $(k_1,k_2)=(0,\pi)$; for $4<|m/r|<6$ there exists a single RH mode centered at $(k_1,k_2)=(\pi,\pi)$.  For $|m/r|>6$ there are no chiral mode solutions. }
\label{fig:DWF2d}
\end{figure}

So the idea is: add a Wilson term, with strength such that the doublers at the corners of the Brillouin zone have $ |\Delta m(k)|$ too large to support a zeromode solution. 
Under separation of variables, one looks for zeromode solutions with $\Psi(x,s) = e^{ipx}\phi_\pm(s)\psi_\pm$ with $\Gamma\psi_{\pm} = \pm\psi$.  One then finds (for $r=1$)
\beq
\pslash_4 \psi_\pm = 0\ ,\qquad -\phi_\pm(s\mp 1) +(m_\text{eff}(s) +1)\phi_\pm(s) =0\ \ ,
\eqn{dwfs}\eeq
where
\beq
m_\text{eff}(s) = m\epsilon(s) + \sum_\mu (1-\cos p_\mu)\equiv m\epsilon(s)  +F(p)\ .
\eeq
Solutions of the form $\phi_\pm(s) = z_\pm^s$ are found with 
\beq
z_\pm = (1+m_\text{eff}(s))^{\mp 1} = (1+m\epsilon(s)  +F(p))^{\mp 1}\ ;
\eeq
they are normalizable if $|z|^{\epsilon(s)}<1$.  Solutions are found for $\psi_+$ only, and then provided that $m$ is in the range  $F(p)<m<F(p)+2$.  (For $r\ne 1$, this region is found by replacing $m\to m/r$.)  However, even though the solution is only found for $\psi_+$, the chirality of the solutions will alternate with corners of the Brillouin zone, just as we found for naive fermions, \eq{altchi}.  The picture for the spectrum in 2d is shown in Fig.~\ref{fig:DWF2d}.  It was first shown in \cite{Kaplan:1992bt} that doublers could be eliminated for domain wall fermions on the lattice;  the rich spectrum in Fig.~\ref{fig:DWF2d} was worked out in \cite{Jansen:1992tw}, where for 4d they found the 
 number of zeromode solutions to be the Pascal numbers $(1,4,6,4,1)$ with alternating chirality,  the critical values for $|m/r|$ being $0,2,\ldots,10$.   One implication of their work is that the Chern Simons currents must also change discontinuously on the lattice at these critical values of $|m|/r$; indeed that is the case, and the lattice version of the Callan-Harvey mechanism was verified analytically in \cite{Golterman:1992ub}.

Fig.~\ref{fig:DWF2d} suggests that chiral fermions will exist in two spacetime dimensions so long as $0<|m/r|<6$, with critical points at $|m/r| = 0,2,\ldots,6$ where the numbers of massless flavors and their chiralities change discontinuously.  In four spacetime dimensions a similar calculation leads to chiral fermions for $0<|m/r|<10$ with critical points at $|m/r| = 0,2,\ldots,10$.  However, this reasoning ignores the gauge fields.  In perturbation theory one would expect the bulk fermions to obtain a radiative mass correction of size $\delta m \sim O(\alpha)$ in lattice units, independent of the extra dimension $s$.  Extrapolating shamelessly to strong coupling, one then expects the domain wall form of the mass to be ruined when $\alpha \sim 1$ for $|m/r| \sim (2n+1)$, $n=0,\ldots,4$ causing a loss of chiral symmetry; near the critical points in $|m/r|$ the critical gauge coupling which destroys chiral symmetry will be smaller.   

While qualitatively correct, this argument ignores the discrete nature of the  lattice.   On the lattice, the exponential suppression $m_\text{res}\sim \text{exp}(-2m s_0)$ found in \eq{mres} is replaced by $\hat T^{L_s} = \text{exp}(-L_s \hat h)$, where $\hat T$ is a transfer matrix in the fifth dimension which is represented by $L_s$  lattice sites.  Good chiral symmetry is attained when $\hat h$ exhibits a ``mass gap", i.e. when all its eigenvalues are positive and bounded away from zero.  However one finds that at strong coupling, rough gauge fields can appear which give rise to near zero-modes of $\hat h$, destroying chiral symmetry, with $m_\text{res}\propto 1/L_s$ \cite{Christ:2005xh,Antonio:2008zz}.  To avoid this problem, one needs to work at weaker coupling and with an improved gauge action which suppresses the appearance of rough gauge fields.

\begin{figure}[t]
\begin{center}
  \includegraphics[width=0.85\textwidth]{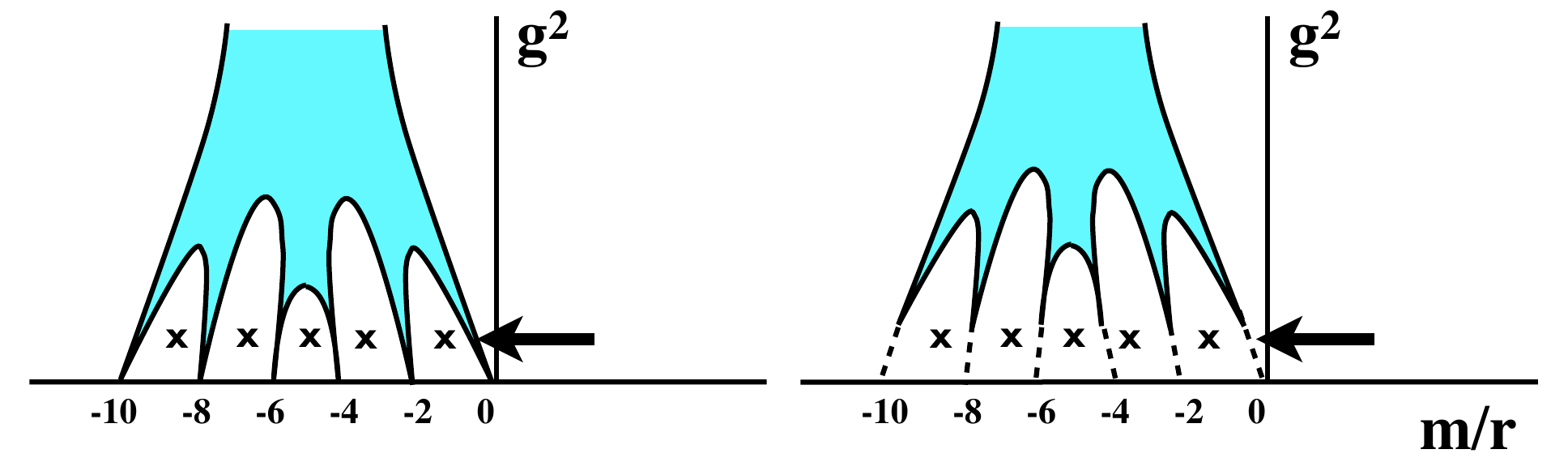}
 \end{center}
\caption{\it A sketch of the possible phase structure of QCD with Wilson fermions  
where the shaded region is the Aoki phase --- pictured extending to the continuum limit (left) or not (right).  When using Wilson fermions one attempts to tune the fermion mass to the phase boundary (arrow) to obtain massless pions; this is only possible in the continuum limit of the picture on the left is correct.  For domain wall fermions chiral symmetry results at infinite $L_s$ when one simulates in any of the regions marked with an ``X". There are six ``fingers" in this picture instead of five due to the discretization of the fifth dimension.}
\label{fig:Aoki}
\end{figure}

At finite lattice spacing the phase diagram is expected to look something like in Fig.~\ref{fig:Aoki} where I have plotted $m$ versus $g^2$, the strong coupling constant.  On this diagram, $g^2\to 0$ is the continuum limit.  Domain wall fermions do not require fine tuning so long as the mass is in one of the regions marked by an ``X", which yield $\{1,4,6,4,1\}$ chiral flavors from left to right.   The shaded region is a phase called the Aoki phase \cite{Aoki:1983qi}; it is presently unclear whether the phase extends to the continuum limit (left side of Fig.~\ref{fig:Aoki}) or not (right side) \cite{Golterman:2005ie}.  In either case, the black arrow indicates how for  Wilson fermions on tunes the mass from the right to the boundary of the Aoki phase to obtain massless pions and chiral symmetry; if the Aoki phase extends down to $g^2=0$ than the Wilson program will work in the continuum limit, but not if the RH side of Fig.~\ref{fig:Aoki} pertains.      See \cite{Golterman:2000zy,Golterman:2003qe} for a sophisticated discussion of the physics behind this diagram.


%

Of course, in the real world we do not see exact chiral symmetry, since quarks and leptons do have mass.  
A mass for the domain wall fermion can be included as a coupling between the LH mode at $s=1$, and the RH mode at $s=N_s$:
\beq
m_q\left[ \mybar\psi(\bfx,1) P_+\psi(\bfx,N_s) + \mybar\psi(\bfx,N_s) P_-\psi(\bfx,1)\right]
\eeq
and correlation functions are measured by sewing together propagators from one boundary to itself for chiral symmetry preserving operators, or from one boundary to the other for operators involving a chiral flip.  The latter will require insertions of the mass operator above to be nonzero (assuming  a neglible $m_\text{res}$) --- just like it should be in the continuum.

 \subsection{Shamir's formulation}

Domain wall fermions are used by a number of lattice collaborations these days, using the formulation of Shamir \cite{Shamir:1993zy,Furman:1994ky}, which is equivalent to the continuum version of domain wall fermions on a slab described above.  The lattice action is given by:
\beq
\sum_{b=1}^5 \sum_\bfx \sum_{s=1}^{N_s}\left[\half\mybar\psi \gamma_b ({\partial_b }^*+ \partial_b)\psi - m\mybar\psi\psi -\frac{r}{2} \mybar\psi {\partial_b }^*\partial_b\psi\right]
\eeq
where the lattice coordinate on the 5d lattice is $\bfn = \{\bfx,s\}$, $\bfx$ and $s$ being  the 4d and fifth dimension lattice coordinates respectively.  The difference operators are 
\beq
\partial_b \psi(\bfn) = \psi(\bfn +\hat \mu_b ) - \psi(\bfn)\ ,\qquad{ \partial_b}^* \psi(\bfn) =  \psi(\bfn)- \psi(\bfn -\hat \mu_b) 
\eeq
where $\hat\mu_b$ is a unit vector in the $x_b$ direction.  In practice of course, these derivatives are gauged in the usual way by inserting gauge link variables.  The boundary conditions are defined by setting fields to zero on sites with $s=0$ and $s=N_s+1$.  I have reversed the sign of $m$ and $r$ from Shamir's original paper, since the above sign for $r$ appears to be relatively standard now.  For domain wall fermions, $m$ has the opposite sign from standard Wilson fermions, which is physics, not convention. The above action gives rise to a RH chiral mode bound to the $s=1$ boundary of the lattice, and a LH chiral mode bound at the $s=N_s$ boundary.

\subsection{The utility of domain wall fermions}

Theoretically, chiral symmetry can be as good a symmetry as one desires if one is close enough to the continuum limit (to avoid delocalization of the zeromode due to large gauge field fluctuations) and large extra dimension.  In practical simulations, the question is whether the residual mass term can be small enough to warrant the simulation cost.  This was reviewed dispassionately and at length in \cite{Sharpe:2007yd}, and I refer you to that article if you are interested in finding out the details.  Currenty, domain wall fermions are being extensively applied to QCD; for some diverse examples from the past year see \cite{Yamazaki:2008hg,Chiu:2008jq,Gavai:2008ea,Torok:2009dg,Ohta:2009uy,Cheng:2009be,Chiu:2009wh}, and  a recent overview \cite{Jansen:2008vs}.  Another recent application has been  to $N=1$ supersymmetric Yang-Mills theory \cite{Giedt:2008cd,Giedt:2008xm,Endres:2008tz,Endres:2009yp,Endres:2009pu}  based on a domain wall formulation for Majorana fermions \cite{Kaplan:1999jn} and earlier numerical work \cite{Fleming:2000fa}.

%
%

\chapter{Overlap fermions and the Ginsparg-Wilson equation}

\section{Overlap fermions }
We have seen that the low energy limit of a domain wall fermion in the limit of large extra dimension is a single massless Dirac fermion, enjoying the full extent of the chiral symmetry belonging to massless fermions in the continuum.  In this low energy limit, the effective theory is four-dimensional if the original domain wall fermion lived in five dimensions.  One might wonder whether one could dispense with the whole machinery of the extra dimension and simply write down the low energy four-dimensional theory to start with.  Furthermore, one would like a four dimensional formulation with exact chiral symmetry, which could only occur for domain wall fermions with infinite extent in the time direction, which is not very practical numerically!

Neuberger and Narayanan found an extremely clever way to  do this, leading to the four-dimensional ``overlap operator" which describes lattice fermions with perfect chiral symmetry.  The starting point is to consider a five dimensional fermion in the continuum with a single domain wall, and to consider the fifth dimension to be time (after all, it makes no difference in Euclidian space). Then $\gamma_5( \Dslash_4+ m(s))$ looks like the Hamiltonian, where $s$ is the new time coordinate, and $m(-\infty)=-m_1$, $m(\infty) = +m_2$, where $m_{1,2}>0$.  The path integral projects onto ground states, and so the partition function for this system is $Z=\braket{\Omega,-m_1}{\Omega,+m_2}$, where the state $\ket{0,m}$ is the ground state of $\CH_4(m)=\gamma_5(\Dslash_4 +m)$  We know that this should describe a massless Weyl fermion.  Note that the partition function is in general complex with an ill-defined phase (we can redefine the phase of $\ket{\Omega,-m_1}$ and $\ket{\Omega,m_2}$ separately and arbitrarily).  If we now instead imagine that the fermion mass function $m(s)$ exhibits a wall-antiwall pair, with the two defects separated infinitely far apart, we recognize a system that will have a massless Dirac fermion in the spectrum, and $Z=\big\lvert \braket{\Omega,-m_1}{\Omega,+m_2}\bigr\vert^2$, which is real, positive, and independent on how we chose the phase for the groundstates.

%
%
We can immediately transcribe this to the lattice, where we replace $\Dslash_4$ with  the four dimensional Wilson operator,
\beq
\CH(m) = \gamma_5 (D_w+m)= \gamma_5 \left(D_\mu\gamma_\mu -\frac{r}{2} D_\mu^2 +m\right)
\eeq
with $D_\mu$ being the symmetric covariant derivative on the lattice, and $D_\mu^2$ being the covariant lattice Laplacian.  Note that $\CH(m)$ is Hermitian, and so its eigenvalues are real.  Furthermore, one can show that it has equal numbers of positive and negative eigenvalues.

We can account for the $\gamma$-matrix structure of $\CH(m)$ explicitly in a chiral basis where $\gamma_5 = \sigma_3\otimes 1$:
\beq
\CH(m) = \begin{pmatrix}B+m & C\cr C^\dagger & -B-m \end{pmatrix}
\eqn{ch4}\eeq
where $B=-\frac{r}{2}\nabla^2$ is the Wilson operator and  $C=D_\mu\sigma_\mu$ where $\sigma_\mu = \{i,\vec\sigma\}$.  For simplicity for $\bra{\Omega,m_2}$ one can take $m_2\to\infty$, in which case $\CH\sim +m_2 \gamma_5$.

We know that $Z=\big\lvert \braket{\Omega,-m_1}{\Omega,+m_2}\bigr\vert^2$ will represent a massless Dirac fermion on the lattice, so long as $0<m_1<2r$, with $m_2$ arbitrary. The groundstates of interest may be written as Slater determinants of all the one-particle wave functions with negative energy.  Let us designate the  one-particle energy eigenstates of $\CH(-m_1)$ and $\CH(m_2)$ to be    $\ket{n,-m_1}$ and $\ket{n,m_2}$ respectively, with 
\beq
\braket{n,m_2}{n',-m_1}\equiv U_{nn'} = \begin{pmatrix}\alpha & \beta \cr \gamma & \delta\end{pmatrix}_{nn'} \ ,\qquad U^\dagger U=1\ ,
\eeq
where the block structure of $U$ is in the same $\gamma$-matrix space that we introduced in writing $\CH$ in block form, \eq{ch4}.
 Now, we want to only fill negative energy eigenstates, so it is convenient to introduce the sign function 
\beq
\varepsilon(\lambda)\equiv  \frac{\lambda}{\sqrt{\lambda^\dagger\lambda}}\ .
\eeq
With $m_2\to\infty$ we have
\beq
\varepsilon(\CH(m_2)) \xrightarrow[m_2\to\infty]{}\gamma_5 = \begin{pmatrix} 1& 0\cr 0 &-1\end{pmatrix}\ .
\eeq
Assuming $\CH(-m_1) $ has no exact zeromodes then, it follows that all eigenvalues  come in $\pm$ pairs (just like the operator $\gamma_5$) and we can choose our basis $\ket{n,-m_1}$ so that 
\beq
\varepsilon(\CH(m_2)) =U\gamma_5 U^\dagger=U \begin{pmatrix} 1& 0\cr 0 &-1\end{pmatrix} U^\dagger\ .
\eqn{epsdef}\eeq
Therefore the Slater determinant we want is
\beq
Z &=& \big\lvert \braket{\Omega,m_2}{\Omega,-m_1}\bigr\vert^2\cr
&=&\big\lvert \det U_{22}\bigr\vert^2\cr
&=&\det\delta^\dagger\det\delta\cr&&\cr
&=& \det\left(\frac{1+\gamma_5\varepsilon(\CH(-m_1))}{2}\right)\ .
\eqn{ol}
\eeq
Some steps have been omitted from this derivation  \cite{Narayanan:2000er}; see exercise \ref{ex4b}.

\begin{exercisebn}
\label{ex4a}
Prove the assertion that if $\CH(-m_1)$ has no zeromodes, it has equal numbers of positive and negative eigenvalues.
\end{exercisebn}
\begin{exercisenb}
\label{ex4b}
You should prove the last step in \eq{ol}, breaking it down to the following steps:
\begin{enumerate} 
\item[(a)] Show that $\det \delta^\dagger=\det \alpha\det U^\dagger $;
\item[(b)] ...so  that $\det\delta^\dagger\det\delta =\det \delta \det \alpha \det U^\dagger  =\det\left[\frac{1}{2} (U+\gamma_5 U\gamma_5)U^\dagger\right]$;
\item[(c)] ...which combines with   \eq{epsdef} to yield \eq{ol}.
\end{enumerate}
 \end{exercisenb}

On the other hand, $Z\propto\det D$, where $D$ is the fermion operator.  So we arrive at the overlap operator (dropping the subscript from $m_1$):
\beq
D &=&1+\gamma_5 \varepsilon(\CH(-m))\cr&&\cr
&=& 1+\gamma_5\frac{\CH(-m)}{\sqrt{\CH(-m)^2}}\cr &&\cr
&=&1+\frac{D_w-m}{\sqrt{(D_w-m)^\dagger (D_w-m)}}\ .
\eqn{olo}\eeq
a  remarkable result.  It was subsequently shown explicitly that this fermion operator can be derived directly from lattice domain wall fermions at infinite wall separation \cite{Neuberger:1997bg,Kikukawa:1999sy}.  Recall from our discussion of domain wall fermions that at least for weak gauge fields, we need $0< m<2r$ in order to obtain one flavor of massless Dirac fermion (where I have set the lattice spacing $a=1$).

\subsection{Eigenvalues of the overlap operator}
Recall that the eigenvalues of the Dirac operator in the continuum  are $\pm i\lambda_n$  for real nonzero $\lambda_n$, plus $n_+$  RH and $n_-$ LH zero modes, where the difference is constrained by the index theorem to equal the topological winding number of the gauge field.  Thus the spectrum looks like a line on the imaginary axis.  What does the spectrum of the overlap operator look like?  Consider
\beq
(D-1)^\dagger (D-1) = \epsilon(\CH)^2 = 1\ .
\eqn{circ}\eeq
Thus $(D-1)$ is a unitary matrix and the eigenvalues of $D$ are constrained to lie on a circle of unit radius in the complex plane, with the center of the circle at $z=1$.  If you put the lattice spacing back into the problem, $D\to a D$ in the above expression to get the dimensions right, and so the eigenvalues sit on a circle of radius $1/a$ centered at $1/a$.  Thus, as $a\to 0$ the circle gets bigger, and the eigenvalues with small magnitude almost lie on the imaginary axis, like the continuum eigenvalues. See the problem below, where you are to show that the eigenfunctions of $D$ with real eigenvalue are chiral.

\subsection{Locality of the overlap operator}
If just presented with the overlap operator \eq{ol} without knowing how it was derived, one might worry that its unusual structure could entail momentum space singularities  corresponding to unacceptable nonlocal behavior in coordinate space.  (From its derivation from domain wall fermions this would be very surprising for sufficiently weakly coupled gauge fields, since the domain wall theory looks well defined and local with a mass gap.)  The locality of the overlap operator (i.e. that it falls off exponentially in coordinate space) was proven analytically in \cite{Hernandez:1998et}, under the assumption of sufficiently smooth gauge link variables, namely that $|1-U|<1/30$. They also claimed numerical evidence for locality that was less restrictive.

\subsection{The value of $m$ and the number of fermions}

For domain wall fermions we found the interesting phase structure as a function of $m/r$, where in the intervals between the critical values $m/r=\{0,-2,-4\ldots,-2(2k+1)\}$ there were $\{1, 2k,\ldots,1\}$ copies of chiral fermions with alternating chirality, as shown in Fig.~\ref{fig:DWF2d}.  One would expect something analogous then for the overlap operator, since it is equivalent to a   domain wall fermion on a $2k$-dimensional lattice with infinite continuous extra dimension.  The equation of motion for the domain wall modes is slightly different than found in \eq{dwfs} due to the continuous dimension:
\beq
\pslash_{2k} \psi_\pm = 0\ ,\qquad \pm\phi_\pm'(s) +(m_\text{eff}(s) +1)\phi_\pm(s) =0\ \ ,
\eqn{dwfsOL}\eeq
where
\beq
m_\text{eff}(s) = m\epsilon(s) + r \sum_\mu (1-\cos p_\mu)\equiv m\epsilon(s)  +r F(p)\ .
\eeq
Solutions are of the form $\phi_\pm(s) = e^{\mp \int^s m_\text{eff}(t) \,dt}$.  With $r F(p)>0$, $\phi_-$ is never normalizable, while for $\phi_+$,   normalizability requires $m/r > F(p)$.  Thus the overlap operator \eq{olo} represents $\{1,1+2k,\ldots,2^{2k}\}$ massless Dirac fermions for $m/r$ in the intervals $(0,2),(2,4),\ldots,(4k,\infty)$. In $2k=4$ dimensions, these flavor numbers are $\{1,5,11,15,16\}$.

\subsection{Simulating the overlap operator}
The overlap operator has exact chiral symmetry, in the sense that it is an exact solution to the Ginsparg Wilson relation, which cannot be said for domain wall fermions at finite $N_s$; furthermore, it is a four-dimensional operator, which would seem to be easier to simulate than a 5d theory.  However, the inverse square root of an operator is expensive to compute, and requires some approximations.  The algorithms for computing it are described in detail in an excellent review by A. Kennedy \cite{Kennedy:2006ax}.  Amusingly, he explains that the method for computing the overlap operator can be viewed as simulating a five-dimensional theory, albeit one with more general structure than the domain wall theory.  For a recent review comparing the computational costs of  different lattice fermions, see the recent review \cite{Jansen:2008vs}.

\begin{exercisebn}
\label{ex4c}
Show that the overlap operator in \eq{olo} has the following properties:
\begin{enumerate}
\item[(a)] At zero gauge field and acting on long wavelength fermion modes, $D\simeq \dslash_4$, the ordinary Dirac operator for a massless fermion.
\item[(b)] It  satisfies the Ginsparg-Wilson equation, \eq{gw1}:
\beq
\{\gamma_5,D\} = D\gamma_5 D\ .
\eeq
\end{enumerate}
\end{exercisebn}
\begin{exercisenb}
\ 
\label{ex4d}
\begin{enumerate}

\item[(a)] Show that one can write $D=1+V$ where $V^\dagger V=1$, and that therefore $D$ can be diagonalized by a unitary transformation, with its eigenvalues lying on the circle $z=1+e^{i\phi}$.
\item[(b)]  Show that, despite $D$ being non-hermitian, normalized eigenstates satisfying $D\ket{z} = z\ket{z}$ with different eigenvalues are orthogonal, satisfying  $\braket{z'}{z} = \delta_{z'z}$
\item[(c)] Show that if $D\ket{z} = z\ket{z}$ then $D^\dagger\ket{z} = z^*\ket{z}$
\item[(d)] Assuming that $\gamma_5 D\gamma_5=D^\dagger$, show that $\expect{z}{\gamma_5}{z}=0$ unless $z=0$ or $z=2$, in which case  $\expect{z}{\gamma_5}{z}=\pm 1$
\end{enumerate}
\end{exercisenb}


\section{The Ginsparg-Wilson equation and its consequences }
In 1982 Paul Ginsparg and Kenneth Wilson wrote a paper about chiral lattice fermions which was immediately almost completely forgotten, accruing  10 citations in the first ten years and none in the subsequent five; today it is marching toward 700 citations.  The reason for this peculiar history is that they wrote down an equation they speculated should be obeyed by a fermion operator in the fixed point action of a theory tuned to the chiral point --- but they did not solve it.  After domain wall and overlap fermions were discovered in the early 1990s, it was realized that they provided a solution to this equation (the domain wall solution only being exact in the limit of infinite extra dimension).  Shortly afterward, M. L\"uscher  elaborated on how the salient features of chirality flowed from the Ginsparg-Wilson equation --- in particular, how anomalies and multiplicative mass renormalization were consequences of the equation, which provided a completely explicit four-dimensional explanation for the success of the overlap and domain wall fermions.

\subsection{Motivation }
A free Wilson fermion with its mass tuned to the critical value describes a chiral fermion in the continuum.  As we have seen, chiral symmetry does not exist on the lattice, but its violation is not evident at low energy, except through correctly reproducing the anomaly.  However, imagine studying this low energy effective theory by repeatedly performing block spin averages.  One would eventually have a lattice theory with all the properties one would desire: chiral fermions and chiral anomalies.  What is the fermion operator in this low energy theory, and how does it realize chiral symmetry?  Motivated by this question, Ginsparg and Wilson performed a somewhat simpler calculation:  they took a continuum theory with chiral symmetry and anomalies, and performed a average of spacetime cells to create a lattice theory, and asked how the chiral symmetry in the original theory was expressed in the resulting lattice theory.

The starting point is   the continuum theory
\beq
Z = \int [d\psi]\,[d\mybar\psi]\, e^{-S(\psi,\mybar\psi)}
\eeq
  I assume there are $N_f$ identical flavors of fermions, and that $S$ is invariant under the full $U(N_f)\times U(N_f)$ chiral symmetry.  We define $\psi_\bfn$ to be localized averages of $\psi$,
\beq
\psi_\bfn = \int d^4 x\, \psi(x) f(\bfx-a\bfn)
\eeq
where $ f(\bfx)$ is some function with support in the region of $|\bfx |\lesssim a$.  Then up to an irrelevant normalization, we can rewrite
\beq
Z &=&  \int [d\psi]\,[d\mybar\psi]\, \int\,\prod_\bfn d\chi_\bfn\,d\mybar\chi_\bfn\, e^{-\left[\sum_\bfn \alpha (\mybar\chi_\bfn-\mybar\psi_\bfn) (\chi_\bfn-\psi_\bfn) - S(\psi,\mybar\psi)\right]}\cr &&\cr
&\equiv &  \int\,\prod_\bfn \,d\chi_\bfn\,d\mybar\chi_\bfn\, e^{- S_\text{lat}(\mybar\chi_\bfn,\chi_\bfn)} \equiv e^{-\mybar\chi D \chi}\ ,
\eeq
where  $\alpha$ is a dimensionful parameter, where $D$ is the resulting lattice fermion operator.   Since there are $N_f$ copies of all the fields, the operator $D$ is invariant under the vector $U(N_f)$ symmetry, so that if $T$ is a $U(N_f)$ generator, $[T,D]=0$.  The lattice action is therefore defined as
\beq
 e^{-\mybar\chi D \chi}= \int [d\psi]\,[d\mybar\psi]\,e^{-\left[\sum_\bfn \,\alpha\,(\mybar\chi_\bfn-\mybar\psi_\bfn) (\chi_\bfn-\psi_\bfn) - S(\psi,\mybar\psi)\right]} \ ,
\eeq
 Note that explicit chiral symmetry breaking has crept into our definition of $S_\text{lat}$ through the fermion bilinear we have introduced in the Gaussian in order to change variables. 

Now consider a chiral transformation on the lattice variables, $\chi_\bfn\to e^{i\epsilon\gamma_5 T}\chi_\bfn$, $ \mybar \chi_\bfn\to \mybar\chi_\bfn e^{i\epsilon\gamma_5 T}$, where $T$   is a generator for a $U(N_f)$ flavor transformation.  This is accompanied  by a corresponding change of integration variables $\psi$, $\mybar\psi$:
\beq
e^{-\mybar\chi e^{i\epsilon\gamma_5 T}D e^{i\epsilon\gamma_5T}\chi} &= &
 \int [d\psi]\,[d\mybar\psi]\, e^{i \int \epsilon \CA\,\Tr T} \,
 e^{-\left[\sum_\bfn \,\alpha\, (\mybar\chi_\bfn-\mybar\psi_\bfn ) e^{2i\epsilon\gamma_5T}(\chi_\bfn- \psi_\bfn) - S(\psi,\mybar\psi)\right]} .\cr &&
\eeq
where $\CA$ is the anomaly due to the non-invariance of the measure $ [d\psi]\,[d\mybar\psi]$ as computed by Fujikawa \cite{Fujikawa:1979ay}:
\beq
\CA = \frac{1}{16\pi^2} \epsilon_{\epsilon\beta\gamma\delta}\Tr F_{\alpha\beta} F_{\gamma\delta}\, 
\eeq
with 
\beq
\int \CA = 2\nu\,\ ,
\eeq
$\nu$ being the topological charge of the gauge field.

Expanding to linear order in $\epsilon$ gives
\beq
-\mybar\chi\{\gamma_5,D\}T\chi \,  e^{-\mybar\chi D \chi}&=& \int [d\psi]\,[d\mybar\psi]\,\left(2\nu\,\Tr T -\sum_\bfn\left[(\mybar\chi_\bfn-\mybar\psi_\bfn )2\alpha\gamma_5 T(\chi_\bfn- \psi_\bfn)\right]\right)\cr &&\qquad \times\, \text{exp}{\left[-\alpha\sum_\bfm (\mybar\chi_\bfm-\mybar\psi_\bfm) (\chi_\bfm-\psi_\bfm) - S(\psi,\mybar\psi)\right]} \cr &&\cr
&=&\left(2\nu\,\Tr T +  \sum_\bfn \,\frac{2}{\alpha} \frac{\delta\ }{\delta\chi_\bfn} \gamma_5T\frac{\delta\ }{\delta\mybar\chi_\bfn} \right)e^{-\mybar\chi D \chi}\cr &&\cr
&=&\left(2\nu\,\Tr T -\frac{2}{\alpha}\Tr \gamma_5 D T-\frac{2}{\alpha} \mybar\chi D\gamma_5 DT\chi \right) e^{-\mybar\chi D \chi}\ .
\eeq
Where no subscript $\bfn $ appears on the $\chi$, its existence and a sum $\sum_\bfn \,$ is implied.  Defining $\alpha \equiv 2/a$ this yields the equation
\beq
\mybar\chi\left(\{\gamma_5,D\}T-a D\gamma_5 DT\right)\chi = \left(a\Tr \gamma_5 D T-2\nu\,\Tr T\right)
\eeq
which must hold for all fields $\chi_\bfn$ and $U(N_f)$ generators $T$.  As a result, we must have $D$ obey
 the Ginsparg-Wilson equation:
\beq
\{\gamma_5,D\}=a\,D\gamma_5 D\ .
\eqn{GW}
\eeq
as well as the relation
\beq
a\Tr \gamma_5 D=2 N_f\nu\ .
\eqn{dnu}
\eeq
This latter equation was not derived in the original Ginsparg-Wilson paper: one can show that $a\Tr \gamma_5 D = -2\,\text{Index}(D) = 2(n_--n_+)$, where  $n_\pm$ are the number of $\pm$ chirality zeromodes of $D$ (see for example, \cite{Luscher:1998pqa}, thereby reproducing  \eq{indexnf}, relating the existence of fermion zeromodes to the winding number $\nu$ in the gauge field. 

Note that the GW relation \eq{GW} is the same equation satisfied  by the overlap operator \cite{Neuberger:1998wv} --- and therefore by the domain wall propagator at infinite wall separation on the lattice, being equivalent as shown in \cite{Neuberger:1997fp,Neuberger:1997bg} ---  as well as by the infinitely separated domain wall propagator  in the continuum \cite{Luscher:2000hn}.  In fact, the general overlap operator derived by Neuberger
\beq
D=1+\gamma_5\epsilon(\CH)
\eqn{olo2}\eeq
is the only explicit solution to the GW equations that is known.


\subsection{Exact lattice chiral symmetry}
Missing from the discussion so far is how the overlap operator is able to ensure multiplicative renormalization of fermion masses (and similarly, multiplicative renormalization of pion masses).  In the continuum, both phenomena follow from the fact that fermion masses are the only operators breaking an otherwise good symmetry.  The GW relation states exactly how chiral symmetry is broken on the lattice, but does not specify a symmetry that {\it is} exact on the lattice and capable of protecting fermion masses from additive renormalization.

L\"uscher was able to solve this problem by discovering the GW relation implied the existence of an exact symmetry of the lattice action: $\int \mybar \psi D\psi$ is invariant under the transformation
\beq
\delta\psi = \gamma_5\left(1 + \frac{a}{2} D\right) \psi\ ,\qquad \delta\mybar\psi = \mybar\psi \left(1 - \frac{a}{2} D\right) \gamma_5\ .
\eeq
Note that this becomes ordinary chiral symmetry in the $a\to 0$ limit, and that it is broken explicitly by a mass term for the fermions.

\subsection{Anomaly}

If this symmetry were an exact symmetry of the path integral, we would run afoul of all the arguments we have made so far: it becomes the anomalous $U(1)_A$ symmetry in the continuum, so it cannot be an exact symmetry on the lattice!  The answer is that this lattice chiral transformation is not a symmetry of the measure of the lattice path integral:
\beq
\delta [d\psi][d\mybar\psi]&=&   [d\psi][d\mybar\psi]\left(\Tr \left[ \gamma_5\left(1 + \frac{a}{2} D\right)\right] +\Tr \left[ \left(1 - \frac{a}{2} D\right)\gamma_5\right]\right) 
\cr&&\cr
&=&  [d\psi][d\mybar\psi] \times a\Tr \gamma_5 D\ ,
\eeq
where I used the relation $d\det M/dx = \det[M] \Tr M^{-1} dM/dx$.  Unlike the tricky noninvariance of the fermion measure in the continuum under a $U(1)_A$ transformation -- which only appears when the measure is properly regulated --- here we have a perfectly ordinary integration measure and a  transformation that gives rise to a Jacobean with a nontrivial phase (unless, of course, $\Tr \gamma_5 D=0$).   To make sense, $\Tr \gamma_5 D$ must map into the continuum anomaly...and we have already seen that it does, from \eq{dnu}.  

What remains is to prove the index theorem \cite{Hasenfratz:1998ri,Luscher:1998pqa}, the lattice equivalent of \eq{idex}.  
From exercise \ref{ex4d} it follows that for states $\ket{z}$ satisfying $D\ket{z} = z\ket{z}$
\beq
\Tr\gamma_5 D = \sum_z \expect{z}{\gamma_5 D}{z} = 2N_f(n^{(2)}_+ - n^{(2)}_-)\ ,
\eeq
where $n^{(2)}_\pm$ are the number of positive and negative chirality states with eigenvalue $z=2$.  We also know that
\beq
0=\Tr\gamma_5 =  \sum_z \expect{z}{\gamma_5 }{z} = (n_+ - n_-) + (n^{(2)}_+ - n^{(2)}_-)\ ,
\eeq
where $n_\pm$ are the number of $\pm$ chirality zeromodes at $z=0$.  Therefore we can write
\beq
\Tr\gamma_5 D =2 (n_- - n_+)\ .
\eeq
Substituting into \eq{dnu} we arrive at the lattice index theorem,
\beq
(n_- - n_+)=\nu N_f
\eqn{indexnflat}\eeq
which is equivalent to the continuum result \eq{indexnf}, and provides an interesting definition for the topological charge of a lattice gauge field.
A desirable feature of the overlap operator is the existence of exact zeromode solutions in the presence of topology; it is also a curse for realistic simulations, since the zeromodes make it difficult to sample different global gauge topologies.  And while it cannot matter what the global topology of the Universe is, fixing the topology in a lattice QCD simulation gives rise to spurious effects which only vanish with a power of the volume \cite{Edwards:2001ei}.

\section{Chiral gauge theories: the challenge}

Chiral fermions on the lattice make an interesting story whose, final chapter on chiral gauge theories has barely been begun.  It is a story that is both theoretically amusing and of practical importance, given the big role chiral symmetry plays in the standard model.  I have tried to stress that the understanding of anomalies has been the key to both understanding the puzzling doubling problem and its resolution. In terms of practical application they are more expensive than other fermion formulations, but have advantages when studying physics where chirality plays an important role.  For a recent review comparing different fermion formulations, see \cite{Jansen:2008vs}.

While domain wall and overlap fermions provide a way to represent any global chiral symmetry without fine tuning, it may be possible to attain these symmetries by fine tuning in theories with either staggered or Wilson fermions.  In contrast, there is currently no practical way  to regulate general nonabelian chiral gauge theories on the lattice.  (There has been a lot of papers in this area, however, in the context of domain wall - overlap - Ginsparg-Wilson fermions;  for a necessarily incomplete list of references that gives you a flavor of the work in this direction, see \cite{Kaplan:1992bt,Kaplan:1992sg,Narayanan:1993ss,Narayanan:1994gw,Narayanan:1995ft,Kaplan:1995pe,Luscher:1998du,Aoyama:1999hg,Luscher:1999un,Kikukawa:2000kd,Kikukawa:2001mw,Kadoh:2007xb,Hasenfratz:2007dp} ).  Thus we lack of a nonperturbative regulator for the Standard Model --- but then again, we think perturbation theory suffices for understanding the Standard Model in the real world.  If a solution to putting chiral gauge theories on the lattice proves to be a complicated and not especially enlightening enterprise, then it probably is not worth the effort (unless the LHC finds evidence for a strongly coupled chiral gauge theory!).  However, if there is a compelling and physical route to such theories, that would undoubtedly be very interesting.

Even if eventually a lattice formulation of the Standard Model is achieved, we must be ready to address the sign problem associated with the phase of the fermion determinant in such theories.  A sign problem has for years plagued attempts to compute properties of QCD at finite baryon chemical potential; the same physics is responsible for poor signal/noise ratio experienced when measuring correlators in multi-baryon states.  To date there have not been any solutions which solve this problem.
We can at least take solace in the fact that the sign problems encountered in chiral gauge theories and in QCD at finite baryon density are not independent!  After all, the standard model at fixed nontrivial $SU(2)$ topology with a large winding number can describe a transition from the QCD vacuum to a world full of iron atoms and neutrinos!  

\bibliographystyle{OUPnamed_notitle}
\bibliography{chiral}

\end{document}